\documentclass[fleqn,usenatbib]{mnras}
\usepackage{newtxtext,newtxmath}
\usepackage[]{hyperref}
\hypersetup{breaklinks=true,colorlinks=true,urlcolor=blue, linkcolor=blue, citecolor=blue, bookmarksopen=true}
\usepackage[T1]{fontenc}

\DeclareRobustCommand{\VAN}[3]{#2}
\let\VANthebibliography\thebibliography
\def\thebibliography{\DeclareRobustCommand{\VAN}[3]{##3}\VANthebibliography}

\usepackage{graphicx}	
\usepackage{amsmath}	
\usepackage{verbatim}
\usepackage{soul}
\usepackage{subcaption}

\title[Interacting twin disc systems]{The SPHERE view of three interacting twin disc systems in polarised light}
\author[P. Weber et al.]{
Philipp Weber,$^{1,2,3}$\thanks{Contact e-mail: \href{E-mail: philipppweber@gmail.com}{philipppweber@gmail.com}}
Sebasti{\'a}n P{\'e}rez,$^{1,2,3}$
Greta Guidi,$^4$
Nicol{\'a}s T. Kurtovic,$^{5}$
Alice Zurlo,$^{6,2,7}$
\newauthor
Antonio Garufi,$^{8}$
Paola Pinilla,$^{9,5}$
Satoshi Mayama,$^{10}$
Rob G. van Holstein,$^{11}$
Cornelis P. Dullemond,$^{12}$
\newauthor
Nicol{\'a}s Cuello,$^{13}$
David Principe,$^{14}$
Lucas Cieza,$^{6,2}$
Camilo Gonz{\'a}lez-Ruilova$^{11,6,2}$ and 
Julien Girard$^{15}$
\\
$^{1}$ {Departamento de Física, Universidad de Santiago de Chile, Av. Victor Jara 3659, Santiago.}\\
$^{2}${Millennium Nucleus on Young Exoplanets and their Moons (YEMS), Chile.}\\
$^{3}${Center for Interdisciplinary Research in Astrophysics and Space Exploration (CIRAS), Universidad de Santiago de Chile, Estaci\'on Central, Chile.}\\
$^4${ETH Zurich, Institute for Particle Physics and Astrophysics, Wolfgang-Pauli-Str. 27,CH-8093 Zurich, Switzerland.}\\
$^5${Max-Planck-Institut für Astronomie, Königstuhl 17, 69117, Heidelberg, Germany.}\\
$^6${Núcleo de Astronomía, Facultad de Ingeniería y Ciencias, Universidad Diego Portales, Av. Ejercito 441, Santiago, Chile.}\\
$^7${Escuela de Ingenier\'ia Industrial, Facultad de Ingenier\'ia y Ciencias, Universidad Diego Portales, Av. Ejercito 441, Santiago, Chile.}\\
$^8${INAF, Osservatorio Astrofisico di Arcetri, Largo Enrico Fermi 5,
50125 Firenze, Italy.}\\
$^{9}${Mullard Space Science Laboratory, University College London,
Holmbury St Mary, Dorking, Surrey RH5 6NT, UK.}\\
$^{10}${The Graduate University for Advanced Studies, SOKENDAI, Shonan Village, Hayama, Kanagawa 240-0193, Japan.}\\
$^{11}${European Southern Observatory, Alonso de Cordova 3107, Casilla
19001, Vitacura, Santiago, Chile.}\\
$^{12}${Institute for Theoretical Astrophysics, Zentrum für Astronomie, Heidelberg University, Albert Ueberle Str. 2, 69120 Heidelberg, Germany.}\\
$^{13}${Univ. Grenoble Alpes, CNRS, IPAG / UMR 5274, F-38000 Grenoble, France}\\
$^{14}${MIT Kavli Institute for Astrophysics and Space Research, 77 Massachusetts Avenue, Cambridge, MA 02139, USA.}\\
$^{15}${Space Telescope Science Institute (STScI), 3700 San Martin Dr, Baltimore MD, 21218, USA.}
}
\date{Accepted XXX. Received YYY; in original form ZZZ}

\pubyear{2022}

\begin{document}
\label{firstpage}
\pagerange{\pageref{firstpage}--\pageref{lastpage}}
\maketitle

\begin{abstract}
Dense stellar environments as hosts of ongoing star formation increase the probability of gravitational encounters among stellar systems during the early stages of evolution. Stellar interaction may occur through non-recurring, hyperbolic or parabolic passages (a so-called ‘fly-by'), through secular binary evolution, or through binary capture. In all three scenarios, the strong gravitational perturbation is expected to manifest itself in the disc structures around the individual stars.
Here, we present near-infrared polarised light observations that were taken with the SPHERE/IRDIS instrument of three known interacting twin-disc systems: AS~205, EM* SR~24, and FU~Orionis. 
The scattered light exposes spirals likely caused by the gravitational interaction.
On a larger scale, we observe connecting filaments between the stars.
We analyse their very complex polarised intensity and put particular attention to the presence of multiple light sources in these systems. The local angle of linear polarisation indicates the source whose light dominates the scattering process from the bridging region between the two stars.
Further, we show that the polarised intensity from scattering with multiple relevant light sources results from an incoherent summation of the individuals' contribution. 
This can produce nulls of polarised intensity in an image, as potentially observed in AS~205.
We discuss the geometry and content of the systems by comparing the polarised light observations with other data at similar resolution, namely with ALMA continuum and gas emission.
Collective observational data can constrain the systems' geometry and stellar trajectories, with the important potential to differentiate between dynamical scenarios of stellar interaction.
\end{abstract}




\section{Introduction}
The applicability of the classical idea of star formation -- a single star forming from a spherically-symmetric and isolated prestellar core \citep[][]{Larson1969,Shu1977} -- has been put to question in recent years. One reason certainly is the observation of stellar birth sites as embedded in thin filaments within molecular clouds \citep[][]{Andre2010}, connoting that large scale processes are integrated in this dynamical environment. Further, dedicated surveys of young, accreting systems found that a relevant fraction of prestellar cores evolve into multiple systems rather than single stars \citep[e.g.][]{Chen2013,Tobin2016,Maury2019}, which can partly explain the multiplicity rate of main-sequence stars \citep[e.g.][]{Raghavan2010,Duchene2013}.

Many concepts of protoplanetary disc dynamics and planet formation were developed in regard of the considerably more quiescent environment of a single stellar host. Yet, planets are also abundant in multiple systems, in circumstellar or circumbinary configuration \citep[][and references therein]{Martin2018}, motivating to further investigate these environments.

There are different processes sought to explain the observation of systems of multiplicity. The idea of turbulent fragmentation predicts that within a gravitationally bound prestellar core or filament, turbulence can create multiple density enhancements that collapse locally depending on an intricate interplay of rotation, gravity and thermal support \citep[e.g.][]{Hoyle1953,Padoan2002,Offner2010}. Even though they may be born with an initially large separation, the protostellar components would progressively approach each other on a comparatively short time-scale \citep[][]{Offner2010}.

A second pathway to multiples' formation takes place at a later stage of the accreting system's evolution. The hypothesis of disc fragmentation assigns the formation of multiples to gravitationally unstable regions in massive fragmenting, primary discs \citep[e.g.][]{Adams1989,Kratter2006,Krumholz2007}. As the location of the secondary's formation in this scenario is limited by the primary's circumstellar disc, this formation mode is expected to lead to much closer companions in comparison to turbulent fragmentation. Indeed, distances measured in population studies of multiple systems reflect this bi-modal formation \citep[][]{Tobin2016}.

Finally, since many stars are believed to form in dense star-forming environments, the probability of stellar encounters can be relevant, potentially producing so-called stellar fly-bys \citep[e.g.][]{Clarke1993,Pfalzner2003,Munoz2015,Cuello2022}. Those star-star encounters are expected to be quite common in the earliest phase of a system's evolution \citep[][]{Pfalzner2013,Winter2018,Pfalzner2021} and may have different consequences for circumstellar material \citep[][]{Clarke1993,Cuello2019}. Mainly depending on the angle of entry, it may tidally truncate the primary disc \citep[e.g.][]{Breslau2014}, modify its inclination \citep[e.g.][]{Xiang-Gruess2016} or cause self-gravitating fragmentation within \citep[e.g.][]{Thies2010}.
Recently, it has been shown that a disc-penetrating fly-by may be the dynamical cause for so-called FU~Orionis events \citep[][]{Borchert2022}, i.e. transient accretion outbursts that increase the stellar brightness by several orders of magnitude for a period of several years. 
For more information on the occurrence and formation of multiples we refer to the most recent detailed review in \citet{Offner2022}.

Exoplanet statistics allow to estimate the significance of stellar multiplicity for the process of planet formation. Notably, \citet{Moe2021} find that when comparing multiples to single stars, the occurrence rate of planetary companions is suppressed for binaries with a stellar separation $a_{\rm bin}\lesssim200\,$au, while no significant difference is found above this value.
This seems to be connected to the presence and lifetime of solids around stars in the protoplanetary phase, as found for tight ($a_{\rm bin}\lesssim40\,$au) and medium separation binaries \citep{Cieza2009,Kraus2012,Zurlo2020,Zurlo2021MNRAS.501.2305Z}. 
The presence of a massive companion dynamically truncates the disc, it is expected to hinder grain growth and promote fast radial drift of dust particles \citep{Zagaria2021a}, creating more challenging conditions for planet-forming processes.

To understand how planet formation may happen in multiple stellar systems, detailed observations of individual systems in their gas-rich phase are of crucial importance. Several multiple systems have been observed in sufficient resolution to expose circumbinary discs \citep[as in GG~Tauri,][]{Dutrey1994,Keppler2020}, individual circumstellar discs \citep[as in HT~Lup or VLA~1623,][respectively]{Kurtovic2018,Sadavoy2018a} and connecting transient gas bridges and filaments (as in IRAS~16293–2422, \citealp{Pineda2012,Sadavoy2018b}, or in Barnard~59, \citealp{Alves2019}). Further, the tidal forces of binary companions have been observed to invoke clearly distinguishable patterns in the circumprimary discs, such as the two opposing spirals in HD~100453 \citep[][]{Bensity2017,Gonzalez2020}.

Another important feature is the degree of alignment of possible circumbinary or circumstellar discs with the binary's orbital plane. Several observational studies suggest that misalignment in multi-disc systems is not uncommon \citep{Czekala2019}. Studying the binary HK~Tau, \citet{Jensen2014} found from CO(3--2) emission that the accompanying discs are inclined by $60^\circ-68^\circ$ towards each other, meaning that at least one of the discs has to be significantly misaligned towards the binary orbital plane. \citet{Brinch2016} detected a similar misalignment for the putatively much younger binary system IRS~43, where the circumstellar discs are misaligned by >$60^\circ$ towards each other while being surrounded by a larger circumbinary disc. Further reports of interacting circumstellar discs with different orientiations were presented for the two-star systems L1551~NE \citep{Takakuwa2017}, V892 Tau \citep{Long2021}, XZ~Tau \citep{Ichikawa2021}, IRAS~04158+2805 \citep{Ragusa2021} and for the multiple systems UX~Tauri \citep{Menard2020} and Z~Canis~Majoris \citep[][]{Canovas2015A&A...578L...1C,Dong2022}. 

It is currently under debate whether some of the misaligned structures observed to be more dynamically-disrupted are evidence of a stellar fly-by rather than a secularly evolving system. A detailed review on the impact of a fly-by on protoplanetary discs was recently given in \citet{Cuello2022}. The most promising candidates for fly-bys are RW\,Aurigae \citep[][]{Cabrit2006,Dai2015} and the previously mentioned cases of UX~Tauri \citep[][]{Menard2020} and Z~Canis~Majoris \citep[][]{Dong2022}. However, we want to highlight that a stellar fly-by is not the only scenario capable of explaining the misaligned discs. \citet{Offner2010} found that gravitationally bound protostellar pairs formed from the same core through turbulent fragmentation retain a randomly oriented angular momentum during their coupled evolution. In the context of turbulent fragmentation it is, therefore, coherent that also accompanying circumstellar discs may show misalignment towards each other, and towards a possible engulfing circumbinary disc without the direct necessity of a fly-by. Global simulations of turbulent fragmentation conducted in \citet{Kuffmeier2019} showed that also a connecting filament of compressed gas emerges between newly formed protostars in multiple systems.

In this paper, we focus on three systems of known disc-disc interactions (introduced in Sec.~\ref{sec:targets}): two hierarchical triplets, namely AS~205 and SR~24, and one binary system, FU~Orionis. We describe the observational details for the  InfraRed Dual-band Imager and Spectrograph (IRDIS) sub-system of the Spectro-Polarimetric High-contrast Exoplanet REsearch (SPHERE) instrument at the Very Large Telescope (VLT) in Sec.~\ref{sec:observations}. We discuss the significance of multiple light sources for polarised light observations in Sec.~\ref{sec:two_light_sources} and present the new near-infrared (NIR) data taken in $H$-band ($\lambda_{\rm obs}=1.625\,$µm) in Sec.~\ref{sec:results}. In Sec.~\ref{sec:discussion}, we compare the new data to archival data from the Atacama Large Millimeter/Submillimeter Array (ALMA) and discuss the states of the three stellar systems. Finally, in Sec.~\ref{sec:conclusions} we summarise the conclusions of this paper.

\section{Targets}\label{sec:targets}
{In the following section we introduce the three targets discussed in this work. Before, we want to warn against the systematical uncertainties within given distance and mass measurements.}

{For some stars the Gaia solution \citep{GaiaDR3} fails to fit the astrometrical measurements, which is indicated by the so-called "Renormalized Unit Weight Error" ({\tt RUWE}). If ${\tt RUWE}\gtrsim1.4$ the Gaia solution should be treated carefully as the goodness-of-fit indicator shows a low level of confidence \citep[][]{elBadry2021}. This can be caused by unresolved or resolved stellar companions \citep[][respectively]{Stassun2021,Kervella2022}, or by surrounding protoplanetary discs \citep[][]{Fitton2022}. 
All three of the above examples for a high {\tt RUWE} are relevant for the three systems we aim to inspect here. 
In Table~\ref{tab:Gaia}, we list the Gaia measurements for the three targets. We additionally list the {\tt ipd\_gof\_harmonic\_amplitude} parameter, which characterises the anisotropy of the Gaia solution and can indicate marginally resolved doubles: objects with ${\tt igha}>0.1$ and ${\tt RUWE}>1.4$ most likely characterise resolved doubles, which have not been correctly handled yet \citep[][]{Fabricius2021,elBadry2021}.}
\begin{table}
\begin{center}
\caption{{Overview of Gaia measurements.}}
\begin{tabular}{ |c||c|c|c|c| }\label{tab:Gaia}
 {\it Object} & $G$ & $d$ [pc] & {\tt RUWE} & {\tt igha}\\
 \hline
 \hline
  {\it AS~205N} & 12.0 & 132$\pm$1 & 4.0  & 0.07 \\
  {\it AS~205S$^{\dagger}$} &13.2 & 142$\pm$3 & 1.7 & 0.52 \\
  \hline
 {\it SR~24N$^{\dagger}$} & 15.1 & -- & -- & 0.74 \\
 {\it SR~24S} & 14.5 & $100\pm2$ & 8.2 & 0.28 \\
 \hline
 {\it FU~OriN} & 9.3 & $408\pm3$ & 1.1  &  0.03 \\ 
 {\it FU~OriS} & -- & -- & --  & --    \\ 
 \hline
 \hline
\end{tabular}
\end{center}
      \small
      {\bf Notes.} $G$ is the $G$-band magnitude, {\tt RUWE} the "Renormalized Unit Weight Error", and {\tt igha} the {\tt ipd\_gof\_harmonic\_amplitude}. Known binaries are marked with by $\dagger$.
\end{table}

{Further, mass estimates from stellar evolution models may be overestimated in case the star is going through the phase of an irregular outburst. Such bursts can lift the luminosity-to-mass relation from the typically applied tracks (e.g. Hayashi track) of stellar evolution models \citep[][]{Jensen2018}.}
\subsection{AS~205}
AS~205 is a young system \citep[$\sim$0.6 Myr,][]{Andrews2018} between the Upper Sco and $\rho$ Ophiuchus star forming regions. In existing literature it is hence associated with the first \citep[as in][]{Barenfeld2016,Andrews2018} or  the latter \citep{Prato2003,Eisner2005,Kurtovic2018}. \citet{Ghez1993} discovered that AS~205 consists of a primary northern and a secondary southern component (AS~205N and AS~205S, respectively), separated by $1\farcs3$.

{Parallax measurements result in distances of $132\pm1\,$pc (${\tt RUWE}=4.0$) for AS~205N and $142\pm3\,$pc (${\tt RUWE}=1.7$) for AS~205S \citep[][]{GaiaDR3}. The discs around AS~205N and AS~205S show visual evidence of gravitational interaction between the components of the system, i.e. a line-of-sight separation of $10\pm 4\,$pc is implausible \citep[][]{Salyk2014,Kurtovic2018}. The disagreement of the two inferred distances highlights that the Gaia measurements are systematically erroneous for AS~205, also indicated by the high {\tt RUWE} of both components.}

AS~205N is classified as a K5 late-type dwarf \citep{Prato2003,Eisner2005} with a mass of $0.87^{+0.15}_{-0.1}\,M_\odot$ \citep{Andrews2018}\footnote{Stellar masses rely on the assumed model. For AS~205N the mass ranges from 0.87$\,M_\odot$ \citep{Andrews2018} to $1.5\,M_\odot$ \citep{Prato2003}. This leads to strong deviations in mass ratios between AS~205N and AS~205S, from $M_1/M_2=0.2$ \citep{Prato2003} to $M_1/M_2=1.1$ \citep{Eisner2005}.}.
The southern component (AS~205S) was found to be a spectroscopic binary itself \citep[AS~205Sa and AS~205Sb,][]{Eisner2005}, making the system  a hierarchical triple. AS~205Sa and AS~205Sb are classified as K7 and M0 stars, with mass estimates of $0.74\,M_\odot$ and $0.54\,M_\odot$, respectively \citep[obtained from fitting spectroscopic data to a stellar evolution model,][]{Eisner2005}.
{Further, \citet{Eisner2005} measured the relative radial velocity of the two southern components to be $\Delta v_{\rm Sa-Sb}=(17.4\pm1.6)\,{\rm km}\,{\rm s}^{-1}$, yielding an upper limit of $a_{\rm Sa-Sb}\leq(3.5\pm0.3)\,{\rm au}$ for the semi-major axis of the southern binary's orbit. This was further constrained by the limit $a_{\rm Sa-Sb}\leq 2\,{\rm au}$ found from a central cavity in mm-flux \citep[][]{Kurtovic2018}.}
Also for AS~205N signatures of a close-in additional companion were proposed \citep{Almeida2017}. Photometric surveys show periodic signatures in the light curve \citep[][]{Artemenko2010,Percy2010}, consistent with radial velocity variations from multi-epoch spectroscopic data \citep{Almeida2017}. The putative object is expected to orbit on a short period of only $\simeq 25\,$ days, estimated to have a mass of $\geq 19.25\,M_{\rm J}$ and an eccentricity of $\simeq 0.34$ \citep{Almeida2017}.

Using the Submillimeter Array (SMA), \citet{Andrews2007} observed an extended dust disc around AS~205N. \citet{Andrews2009} also detected a reduced, spatially confined signal from AS~205S, pointing towards the presence of a circumbinary disc around the southern component.

The system was prominently part of the ALMA Disk Substructures at High Angular Resolution Project \citep[DSHARP,][]{Andrews2018,Kurtovic2018}, providing more precise information about the dust content.
One critical point in understanding the dynamics of the system is the positioning of the two binary components in three-dimensional space. \citet{Andrews2007} had pointed out that an extended dust disc without strong perturbations around AS~205N suggests that the binary orbit is observed inclined, such that the stars true separation is significantly larger than projected to the sky plane. Yet, \citet{Kurtovic2018} detected signs of tidal interaction: the symmetric continuum emission around the primary is super-imposed by two opposing spirals of low-contrast to the background disc, and the CO moment 0 map shows a gas bridge connecting the two sources \citep[as previously noted, ][]{Salyk2014}. Additionally, \citet{Kurtovic2018} also detected substructure in the small disc around the southern source.

By fitting 2D Gaussians to the discs observed in the continuum, \citet{Kurtovic2018} were able to measure inclination, $i$, and position angle, PA, of the discs separately ($i_{\rm N} =20.1^\circ\pm3.3^\circ$, PA$_{\rm N} =114.0^\circ\pm11.8^\circ$ and $i_{\rm S} =66.3^\circ\pm1.7^\circ$, PA$_{\rm S} =109.6^\circ\pm1.8^\circ$). This indicates a large misalignment between the two discs.

\subsection{SR~24}
EM$^\ast$~SR~24 (from now on SR~24) is a hierarchical multiple system located in the Ophiuchus
star-forming region. It consists of a primary (SR~24S) located to the south and a secondary (SR~24N) located to the north, at a $PA$ of 348$^\circ$ and a projected separation of $5\farcs1$ \citep[][]{GaiaDR3}. The northern component consists of a binary system in itself \citep[SR~24Na and SR~24Nb,][]{Simon1995} that follows an eccentric orbit $e = 0.64^{+0.13}_{-0.10}$ with a period of $111^{+105}_{-33}\,$yr \citep[][]{Schaefer2018}. \citet{Correia2006} inferred the stellar masses as $M_{\rm S}>1.4\,M_\odot$, $M_{\rm Na}=0.61\,M_\odot$ and $M_{\rm Nb}=0.34\,M_\odot$ assuming a distance of $\approx$~$160\,$pc. Given the updated distance of the system of $100\pm2\,$pc \citep[measured towards the southern source,][{\tt RUWE}=8.2]{GaiaDR3}, the masses might be overestimated. The analysis of orbital elements in \citet{Schaefer2018} suggests that the combined mass of the northern binary system is merely 0.48$\pm0.12\,M_\odot$.

SR~24 has been observed at different wavelengths over the past decades. \citet{Nuernberger1998} observed SR~24 with the IRAM 30$\,$m Millimeter Radio Telescope, inferring the presence of cold dust around SR~24S from emission at a wavelength of $1.3\,$mm. The simultaneously proposed lack of cold dust around SR~24N \citep{Nuernberger1998} was contrasted by significant flux detection at $10\,$µm, pointing to the local presence of warm dust \citep{Stanke2000}.

\citet{Mayama2010} presented $H$-band ($\lambda = 1.6\,\mu$m) observations using the Coronagraphic Imager with Adaptive Optics (CIAO) at the Subaru Telescope.
In this article the authors also showed optical data collected by the Hubble Space Telescope (proposal ID 7387, PI: Karl Stapelfeldt). Both optical and infrared scattered light exposed an elongated structure bridging SR~24N and SR~24S, as well as a large scale spiral connected to SR~24S, extending southwards.

SR~24 was observed with ALMA in observation cycle 0 in band 9, showing the  $440\,$µm continuum and $^{12}$CO~(6--5) line emission \citep{vanderMarel2015}, in cycle 2 to study the continuum at $1.3\,$mm (band 6) and molecular emission of $^{13}$CO and C$^{18}$O~($J$=2--1) \citep{Pinilla2017,fernandez-Lopez2017} and in cycle 3 in band 3 at $2.75\,$mm \citep{Pinilla2019}. These data suggest an inclination of $i\approx47^\circ.6$ and a position angle of $PA\approx26^\circ.8$ for the disc around SR~24S \citep[][]{Pinilla2019}. The band 6 data support the claim of \citet{Nuernberger1998} that SR~24N seems to be stripped of evolved, cold dust grains, as also in the ALMA band 6 data the continuum emission is exclusively limited to the southern source. \citet{vanderMarel2015}, however, report 440$\,$µm emission that is too extended to be recovered by their shortest baselines, suggesting contributions from scales $\gtrsim 4\farcs$ distant of SR~24S. This is consistent with relevant contribution of SR~24N at this wavelength.

The data taken in band 3 revealed an inner disc around SR~24S, inside the cavity that was previously reported from band 6 data. \citet{Pinilla2019} linked this detection to dust thermal emission of large grains and promote the idea that a gap-opening embedded planet may be responsible for the observed structures.
\citet{fernandez-Lopez2017}, moreover, suggested based on gas kinematics that the circumprimary and circumsecondary discs are strongly misaligned. This is further supported by recent near-infrared (NIR) polarimetric imaging at the HiCIAO instrument at Subaru Telescope \citep{Mayama2020}.

SR~24 was also part of the `Ophiuchus DIsc Survey Employing ALMA' programme \citep[ODISEA,][]{Cieza2019}. The dust ring around SR~24S was measured to be limited to lie between $\sim 30\,$au and $\sim 58\,$au with a dust mass of $\sim70\,$M$_\oplus$ \citep[][]{Cieza2021}. In \citet{Cieza2019} there is no detection of the inner disk in SR~24S, neither of dust continuum emission around SR~24N.

\subsection{FU~Orionis}
FU~Orionis is the archetype for an eponymous group of objects, characterised by the extreme brightness changes they exhibit on an annual time scale \citep[][]{Herbig1966, Audard2014}.
The extreme brightness variability has been linked to abrupt mass transfer from a surrounding accretion disc onto a young, low-mass T~Tauri star \citep[][]{Hartmann1996}. In this scenario, the accretion triggers a stellar outburst that is visible as a transient increase of brightness.

The system's distance is $408\pm 3\,$pc \citep{GaiaDR3}.
\citet{Wang2004} discovered that the FU~Orionis system is a binary with a northern and a southern component, separated by a projected distance of 0\farcs5 between the stars.

 The mass of the northern star was updated from molecular line observations to $M_{\rm N}=0.6\,M_\odot$ \citep[][]{Perez2020}. The northern source is typically treated as the primary as it is brighter by about 4$\,$mag in the NIR \citep[][]{Beck2012}. However, \citet{Beck2012} proposed that FU~OriS is actually the more massive of the pair with a mass of $M_{\rm S}=1.2\,M_\odot$ as estimated from stellar evolution models.
The enormous brightness contrast between the two objects was attributed to strong line-of-sight extinction for the southern source deduced from strong reddening. While \citet{Beck2012} estimate a flux attenuation of $2-3\,$mag of the full NIR spectral range,  \citet{Pueyo2012} estimate this value to be as high as $8-12\,$mag.

FU~Orionis has been observed in NIR polarised light by \citet{Liu2016} and \citet{Takami2018} using the Subaru/HiCIAO instrument in $H$-band, showing that most of the surrounding material is centred on the northern star. \citet{Laws2020} observed the system in $J$-band using Gemini/GPI to reveal intricate asymmetric substructures.

CO observations conducted with ALMA show an extended emission centred on the location of the northern star \citep[][]{Hales2015,Perez2020}, but interestingly the denser molecular tracer HCO+ is centred on FU~OriS. In the band~6 continuum, \citet{Perez2020} detected extended circumstellar discs for both the northern and southern binary components and inferred similar inclinations and position angles for the two discs ($i_{\rm N} =37.7^\circ\pm0.8^\circ$, PA$_{\rm N} =133.6^\circ\pm1.7^\circ$ and $i_{\rm S} =36.4^\circ\pm1.2^\circ$, PA$_{\rm S} =137.7^\circ\pm3.7^\circ$).

In contrast to AS~205 and SR~24, FU~Orionis is known to be in an active state of a stellar outburst, potentially connected to the observed variable structures. This makes it an interesting object to look at and compare with. 

\section{Observations and data reduction}\label{sec:observations}
\subsection{SPHERE IRDIS/DPI}
All three stellar systems were observed in dual-beam polarimetric imaging mode (DPI, \citealp{deBoer2020,vanHolstein2020}) with the InfraRed Dual-band Imager and Spectrograph (IRDIS, \citealp{Dohlen2008}) at VLT/SPHERE \citep{Beuzit2019}. The observations were carried out in $H$-band ($\lambda=1.625\,$µm), using the {\it N\_ALC\_YJH\_S} coronagraph (185$\,$mas diameter) centred on the northern sources of the {respective} stellar system.

The observing conditions are summarised in Table~\ref{tab:obs}. We observed AS~205 and SR~24 in the night of 19 May 2017 in delegated visitor mode (programme-ID: 099.C-0685(A), PI: S.~Pérez). 
AS~205 yielded very good data with eleven polarimetric cycles and 16$\,$s integration time per frame. We discarded the fifth cycle due to a bad frame resulting in a total on-target time of 10.7$\,$min, completed by several centre, sky and flux measurements.

The observation of SR~24, on the other hand, posed a challenge as the objects constitute the faintest stars observed by SPHERE ($R>14$, \citealp{Wilking2005}). SPHERE’s adaptive optics (AO) system is measured to operate robustly for $R<12$ \citep{Beuzit2019}, which highlights the challenge of observing SR~24. After trying for both stars, we managed to centre the AO on SR~24N, the brighter one of the two objects in $R$-band, and took a single polarimetric cycle of 96$\,$s integration time per frame. The star centre frame is corrupted and we measure the centre of the frame by fitting two-dimensional Gaussians to the intensity maximum. The binary companion SR~24S is just on the edge of the field of view.

FU~Orionis was observed in the night of 17 December 2016 (programme ID: 098.C-0422(B), PI: D.~Principe) with an on-source exposure time of 8$\,$min, in five polarimetric cycles with 24$\,$s per frame.
To precisely track the position of the northern stellar component and account for the flux of both stars we additionally obtained flux and centre frames in all three cases.

\begin{table}
\begin{center}
\caption{{Overview of observations.}}
\begin{tabular}{ |c||c|c|c|c| }\label{tab:obs}
 {\it Object} & date & $t_{\rm exp}$ [min] & seeing ["] & $v_{\rm wind}$ [m/s]\\
 \hline
 \hline
  {\it AS~205} & 2017-05-19 & 10.7 & 0.36--0.61 &  6.5--8.2 \\
 {\it SR~24} & 2017-05-19 & 6.4 & 0.44--0.75 &  6.0--8.2\\
 {\it FU~Ori} & 2016-12-17 & 8.0 & 0.40--0.55  &  1.9--4.8  \\ 
 \hline
 \hline
\end{tabular}
\end{center}
      \small
      {\bf Notes.} {The date of the observation is given in the format year-month-day, $t_{\rm exp}$ is the total on-source exposure time, the seeing is measured by DIMM (Differential Image Motion Monitor), $v_{\rm wind}$ is the wind speeds measured by ASM (astronomical site monitor) at 30$\,$m. Seeing and wind are given by their minimum and maximum value during the observation.}
\end{table}

As the direct light from a star is unpolarised, this light is strongly suppressed in images of polarisation. The DPI mode of IRDIS can, therefore, reach contrasts close to the photon-noise limit \citep[][Appendix E]{vanHolstein2021}, revealing the polarised light scattered off the dusty structures surrounding the star.
The polarisation state of observed light is described by its Stokes vector $ \mathbfit{S} = [I,Q,U,V]^{T}$ \citep{Stokes1851}, with the total intensity $I$, the linear polarisation components $Q$ and $U$ (rotated by $45^\circ$ with respect to each other), and the circular polarisation component $V$. From $Q$ and $U$, the linearly polarised intensity is calculated as:
\begin{equation}\label{equ:PI}
    PI = \sqrt{Q^2 + U^2}\,,
\end{equation}
Alternatively, most recent works make use of the azimuthal Stokes parameters $Q_{\phi}$ and $U_{\phi}$ \citep[][]{Schmid2006,deBoer2020}:
\begin{equation}\label{equ:QUphi}
 \begin{cases}
    Q_{\phi}=-Q\cos{\left(2\phi\right)}-U\sin{\left(2\phi\right)} \\ 
    U_{\phi}=+Q\sin{\left(2\phi\right)}-U\cos{\left(2\phi\right)}
 \end{cases}
\end{equation}
with $\phi$ being the azimuth angle. In the case of single scattering of light originating from the coordinate centre, $Q_{\phi}$ is equal to $PI$ except for the reduced noise due to the lack of squared operation.
{Hence, in systems around single stars, $U_\phi$ can be used to indicate regions where light is scattered more than once before reaching the observer \citep[][]{Canovas2015,Pohl2017}. In a system of multiple light sources, however, $Q_\phi$ neglects to capture large parts of the single-scattering from the off-centred light source. Depending on the position in the disc, the scattering induced by the secondary contributes to the signal in $U_\phi$ and may add a negative component to $Q_\phi$. We discuss a possible analysis of the $Q_\phi$ and $U_\phi$ images under consideration of the two resolved light sources in Appendix~\ref{appendix:QUphi} but refer to the total linearly polarised intensity, $PI$, otherwise.}

Finally, the degree of linear polarisation, $DoLP$, and the angle of linear polarisation, $AoLP$, are calculated as:
\begin{eqnarray}
    {DoLP} &=& \frac{PI}{I}\,,\\
    {AoLP} &=& \frac{1}{2} {\arctan \left( \frac{U}{Q}\right)}\,.\label{equ:AoLP}
\end{eqnarray}

To reduce the SPHERE/IRDIS data, we employ the reduction pipeline IRDAP\footnote{\hyperlink{https://irdap.readthedocs.io}{irdap.readthedocs.io}} (IRDIS Data reduction for Accurate Polarimetry, version 1.3.3, \citealp{vanHolstein2020}). This pipeline uses a detailed model of the SPHERE optical system, and, therefore, allows to directly correct for the instrumental polarisation and polarisation crosstalk without using the data itself.
From the IRDAP pipeline the $Q$ and $U$ products are obtained via the double-differencing method \citep[see e.g.][]{vanHolstein2020}. 

\subsection{Archival ALMA data}\label{subsec:ALMAdata}
Additionally to the NIR polarised data, we re-analyse high-resolution ALMA band 6 data on SR~24 from the ODISEA long-baseline programme \citep[programme ID: 2018.1.00028.S,][]{Cieza2019, Cieza2021}. After running the CASA pipeline script, we applied two rounds of phase self-calibration to the continuum spectral windows, increasing the signal-to-noise of the cleaned images by about 90\%. We produce a global clean image of SR~24S with CASA \texttt{tclean} using robust = 0.5, at a resolution of 0\farcs035 $\times$ 0\farcs030. To look for fainter emission from the northern component, we applied \texttt{tclean} with natural weighting resulting in a beam of 0\farcs060 $\times$ 0\farcs055. Further, we applied a JvM correction \citep[][]{JvM1995,Czekala2021}, i.e. we scale the residual image by the ratio between the volumes of the dirty and clean beam (where we find a factor $\varepsilon = 0.34$), and add it to the model components to obtain the final clean image. A detailed description of the application of the JvM correction can be found in \citet{Czekala2021}. We used the JvM correction with caution as it has been noted that it may spuriously reduce the image residuals \citep{Casassus2022}.
Finally, when inspecting the northern source, we correct the primary beam as the phase centre of the ALMA observations was set on the southern source SR~24S. 

\subsection{Archival NACO data}\label{subsec:NACO}
The stellar magnitudes of AS~205N, SR~24N and SR~24S have been measured within the 2MASS survey \citep{2003yCat.2246....0C}.
In the ESO archive we also found NIR imaging data from VLT/NACO \citep[][]{Lenzen2003,Rousset2003}. 

AS 205 was observed with NACO in $J$, $H$, and $K_{S}$ bands during the night 2004-06-11 under good conditions. The project ID of the observations is 073.C-0121(A). The same target was observed again in the $L'$ filter during the night 2019-06-06, project ID 0103.C-0290(A). The two components are visible in all the filters.
SR~24 was observed with NACO in $L'$ band on 2004-06-26 (073.C-0530(A)).

We reduced this archival NACO data on AS~205 and SR~24 following the procedure presented in \cite{Zurlo2020,Zurlo2021MNRAS.501.2305Z}. We refer the reader to those publications for further details. 
We calculated the contrast of AS~205S with respect to the primary. In $J$, $H$ and $K_S$ bands we thus get absolute magnitudes for AS~205S based on the 2MASS magnitudes for AS~205N \citep{2003yCat.2246....0C}. For both AS~205 and SR~24 we complement the list with a contrast measurement in $L'$-band.

The results from the analysis are listed in Table~\ref{tab:mags}, completed with results for FU~Orionis adapted from \citet{Beck2012}.
In this table, we also provide uncertainties for the magnitude measurements.
In the case of the NACO data, these uncertainties represent a conservative estimate based on the contrast between the two components. 
When applying the magnitude measurements provided in Table~\ref{tab:mags} to other observing epochs, we point out that the temporal variability of the stars' luminosity may be much larger than the provided uncertainties (especially for the case of FU~Orionis).
\begin{table}
\begin{center}
\caption{Apparent brightness of binary components in mag. References: [1] 2MASS \citep{2003yCat.2246....0C}, [2] this paper (NACO data as described in Sec~\ref{subsec:NACO}), [3] \citet{Beck2012}.}
\begin{tabular}{ |c|c|c|c|c }\label{tab:mags}
 {\it Object} & J & H & K$_{\rm s}$ & $\Delta$mag(L) \\
 \hline
 \hline
  {\it AS~205N$^{[1]}$} & ${8.06{\pm0.02}}$ & ${6.75{\pm0.03}}$ & ${5.78{\pm0.02}}$ & —  \\
 {\it AS~205S$^{[2]}$} & ${9.3{\pm0.3}}$ & ${7.8{\pm0.3}}$ & ${6.8{\pm0.3}}$ & ${0.3{\pm0.3}}$ \\
 \hline
 {\it SR~24N$^{[1]}$} & ${10.37{\pm{0.03}}}$ & ${8.64{\pm{0.09}}}$ & ${7.55{\pm{0.05}}}$ &— \\
  &  &  &  & ${-0.3{\pm0.3}}^{[2]}$ \\
 {\it SR~24S$^{[1]}$} & ${9.75{\pm{0.03}}}$ & ${8.17}{\pm{0.04}}$ & ${7.06}{\pm{0.03}}$ & — \\
 \hline
 {\it FU~OriN$^{[3]}$} & ${6.5{\pm0.1}}$ & ${5.7{\pm0.1}}$ & ${5.1{\pm0.1}}$  & — \\ 
 {\it FU~OriS$^{[3]}$} & ${10.5{\pm0.3}}$   & ${9.9{\pm0.2}}$ & ${9.1{\pm0.2}}$ & — \\ 
 \hline
 \hline
\end{tabular}
\end{center}
\end{table}

\section{Scattered light polarisation in multiples}\label{sec:two_light_sources}
In recent years, NIR observational techniques and data reduction have mainly been developed to expose dust structures around a central star. In this paper we analyse three systems that host multiple relevant sources of light. Hence, the interpretation of measured polarised scattered light has to account for resulting effects.
We come to similar conclusions as presented in the appendix of \citet{Keppler2020} for the specific example of GG~Tauri~A.

The state of linear polarisation is described by the $Q$- and $U$-components of the Stokes vector, from which polarised light intensity and angle can be calculated according to equation~(\ref{equ:PI}) and equation~(\ref{equ:AoLP}), respectively.
The resulting Stokes vector of scattering from one single light source, $\mathbfit{S}_{\rm out}$, can be calculated by applying the so-called M\"uller matrix, \mathbfss{M}, to an incoming polarisation state, $\mathbfit{S}_{\rm in}$, i.e.: $\mathbfit{S}_{\rm out} = \mathbfss{M}  \mathbfit{S}_{\rm in}$.
The M\"uller matrix is a $4\times4$-operator describing the modifications applied to a state of radiation due to its interaction with matter \citep{Kliger1990,Tinbergen2005}. For light scattering off dust grains, its elements depend on the wavelength of the light, on the scattering angle (i.e. the angle between incoming and outgoing radiation) and on material specific characteristics (such as grain sizes and compositions).
For a specific location in systems with a single source of light, there is a single scattering angle that traces the radiation from source to observer (assuming that the observed light is dominated by single-scattering events).

In regions around binaries or higher order ($n$) multiples several scattering processes are at play, i.e. there is not only one direction of incident light as it is typically assumed for dust structures around single stars. 
The resulting state of scattered light, therefore, is an intensity superposition of all relevant scattering processes\footnote{\citet{Tinbergen2005} points out that this is only true if the outgoing state is averaged over a time that is considerably longer than the time between different scattering events.}:
\begin{equation}
     \mathbfit{S}_{\rm out} = \sum_{\rm i=1}^{n} \left(\mathbfss{M}  \mathbfit{S}_{\rm in}\right)_{\rm i}=\sum_{\rm i} \mathbfit{S}_{\rm out, i} \,.
\end{equation}
Whereas the total intensity in such a case is simply the linear sum of single scattering events, it becomes evident that the polarised intensity results from an incoherent sum,
\begin{equation}\label{equ:incoherentsum}
    PI = \sqrt{\left(\sum_{\rm i} Q_{\rm i}\right)^2 + \left(\sum_{\rm i} U_{\rm i}\right)^2} \leq \sum_{\rm i} PI_{\rm i} \,.
\end{equation}
Formulated verbally, this means that adding polarised intensities from different sources does not necessarily increase the polarised intensity observed from a certain scattering region. Other than in a single star system, the observed linearly polarised intensity image does not only depend on the incident intensities and the scattering characteristics, but also on the intensities' partition into $Q$- and $U$-components (and thus the angle of linear polarization), and hence on the stars' spatial relation to some location in the image, $\mathbfit{x}$.

\begin{figure}
\vspace{-0.5cm}
\includegraphics[width=\columnwidth]{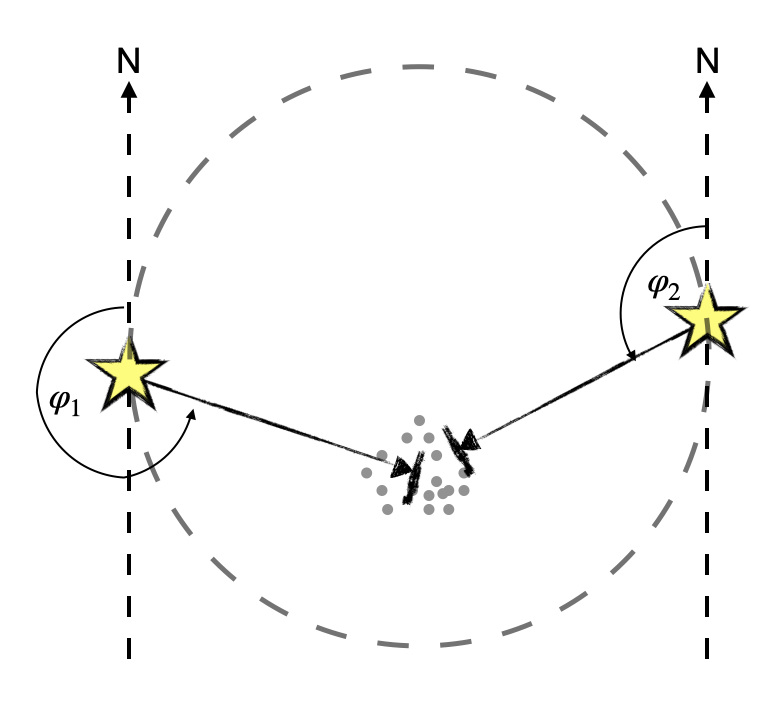}
\caption{Sketch of two stars illuminating a region of dust{, observed as if it were on the plane of the sky, north is up, east is to the left}. The scattered light of each single star is expected to be polarised in the direction perpendicular to the direction towards the star (under the assumption that it is scattered only once). The dashed circle defines the locations where the two angles, $\varphi_1$ and $\varphi_2$, are perpendicular.}
\label{fig:twostars}
\end{figure}

We emphasise that we assume the polarised intensity image to be dominated by single-scattering events, i.e. the photon from a light source is scattered only once before being observed. This results in an angle of linear polarisation that is aligned perpendicular to the line connecting the star and scattering location. 
A simplified case is sketched in Fig.~\ref{fig:twostars}, which shows an observation of scattered light from a dusty region that is illuminated by two stars. 
With the definition of $\varphi_1$ and $\varphi_2$ as the angles towards the image location $\mathbfit{x}$, positive east of north with respect to the stars, we can decompose the linearly polarised intensity produced by scattering of one of the star's radiation, $PI_{\rm i}$, into its directional components:
\begin{eqnarray}
    Q_{\rm i} &= -\, PI_{\rm i}(\mathbfit{x}) \cos (2 \varphi_{\rm i}(\mathbfit{x}))\,, \\
    U_{\rm i} &= -\, PI_{\rm i}(\mathbfit{x}) \sin (2 \varphi_{\rm i}(\mathbfit{x}))\,.
\end{eqnarray}
The $PI$ that would be observed from such a system can then be calculated via equation~({\ref{equ:incoherentsum}}). Inserting the expressions for $Q_{\rm i}$ and $U_{\rm i}$ into equation~(\ref{equ:incoherentsum}) further shows that there is the possibility of vanishing polarised intensity through incoherent summation. The locations where this occurs in a system of two relevant light sources are given by two conditions:
\begin{equation}
    \varphi_1 = \varphi_2\,\pm \frac{\pi}{2}\,, \qquad \text{and} \qquad PI_1=PI_2\,,
\end{equation}
so, where the directions of incident light are perpendicular towards each other, and the individual scattered polarised intensities are equally strong. Here, we expect to observe two nulls in $PI$ images. The condition of orthogonality demands that such points of polarisation cancellation lie on a circle in the image plane centred on the projected centre between the two stars and with the projected distance of the two stars as diameter, sketched as a dashed circle in Fig.~\ref{fig:twostars}.

\section{Results}\label{sec:results}
\begin{figure*}
    \centering
    \includegraphics{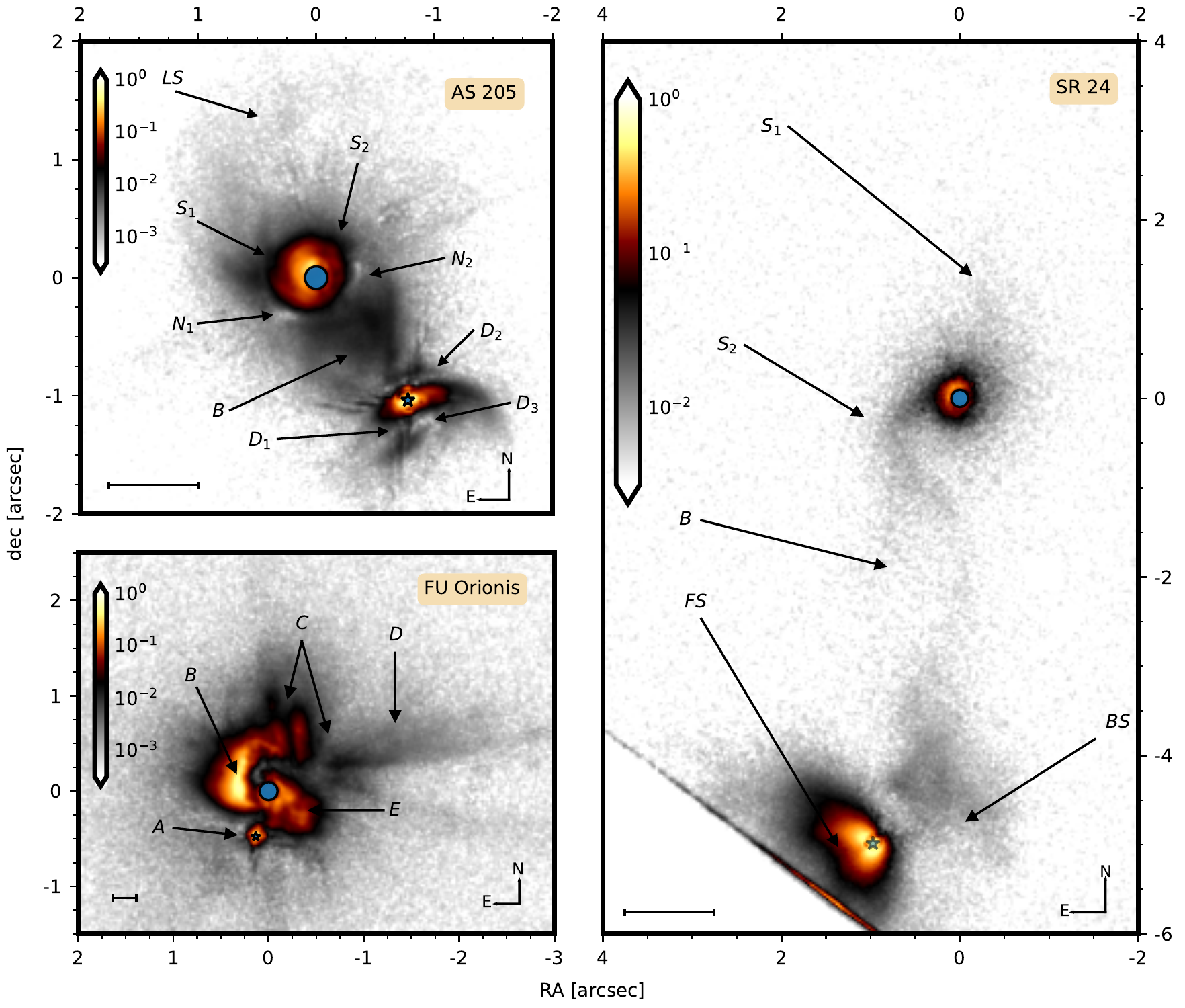}
    \caption{SPHERE/IRDIS observations of three twin disc systems: AS~205, SR~24 and FU~Orionis. North is up, East is to the left. We show the normalised polarised light intensity in logarithmic colour scale with the $1\sigma_{\rm rms}-$level as the lower limit and the maximum intensity as the upper limit. The coordinates are centred on the respective northern source. The central circles mark the coronagraph's diameter. The location of the companions in the images are marked by a star. The length of the bar in the lower left corner of each panel is equivalent to a projected distance of 100$\,$au. Arrows and letters are referred to in the text. For SR~24 the southern source is very close to the edge of the detector which is why the lower frame is visible there.}
    \label{fig:mosaic}
\end{figure*}

In Fig.~\ref{fig:mosaic} we present the linearly polarised light images observed with the SPHERE/IRDIS instrument in $H$-band of the twin-disc systems AS~205, SR~24 and FU~Orionis.
In all three systems we detect structures of polarised flux around both binary components and significant signal from inter-binary, bridging material. We analyse the system-specific structures in the following individual paragraphs.
\subsection{AS~205}
The observation of AS~205 is portrayed in the top-left panel of Fig.~\ref{fig:mosaic}. The coordinates are centred on the northern component, which we measure to be about one order of magnitude brighter in total intensity than the southern companion (see Table~\ref{tab:mags}). The intensity peak of AS~205S is marked as a star in Fig.~\ref{fig:mosaic}, centred at relative coordinates of $(-0\farcs777, -1\farcs036)$ with respect to the northern source. It is slightly off-set from the Gaia-measured separation $(-0\farcs829, -1\farcs050)$. The projected angular separation together with the distance of AS~205 measured by Gaia provides a lower limit of the components deprojected separation of $176.6\,$au.

\subsubsection{Observed Structure}\label{subsubsec:as205_observedstruc}
The scattered light image reveals discs around both AS~205N and AS~205S, in both discs we can distinguish sub-structure and variations of the polarised intensity.
The disc around AS~205N shows two subtly discernible spiral arms, one extending clockwise from the south-west ($S_1$), the other extending clockwise from the north-east ($S_2$). $S_2$ spans about 90$^\circ$ in azimuth and ends abruptly, $S_1$ extends further, almost 180$^\circ$ in azimuth.

The image still shows traces of the diffraction spikes around the location of AS~205S.
The disc around AS~205S is observed under a high inclination \citep[$i = 66.3^\circ$,][]{Kurtovic2018}. The curved structure opening to the south-west suggests a flared disc profile of a vertically-extended disc. This also suggests that the radiation from the bright parts, inner binary and inner regions, is affected by foreground absorption and scattering from (sub-)µm dust grains along the line-of-sight. The disc around AS~205S shows several diminutions in polarised intensity (labelled $D_1$, $D_2$ and $D_3$).

In an inclined disc, changes of the polarised intensity can be due to the angular dependency of the dust grains' scattering phase function, that may promote or hamper the scattering efficiency under different scattering angles \citep[e.g.][]{Perrin2009}. Typically, forward scattering is promoted while backward scattering is hampered \citep[e.g.][]{Min2012,Stolker2016}. In the disc around AS~205N we can see that the north-eastern direction of the disc appears brighter than its south-western counterpart, suggesting the disc's near side directed to the north-east.

{Also a local change in dust-size distribution or dust composition is expected to modify the scattering properties of dust grains and affect the polarised intensity image \citep[][]{Min2012}.
Given the high measured inclination of the disc around AS~205S \citep[$i\sim 66^\circ$,][]{Kurtovic2018} we suggest that the decrement $D_2$ traces disc areas close to the mid-plane that are shielded from stellar irradiation by the surrounding disc.}

{More localised decrements of polarised intensity can be a signal of shadowing from unresolved material close to the light source(s), such as inner discs \citep[][]{Marino2015}, self-shadowing by a spetroscopic binary \citep[][]{DOrazi2019} or inner circumplanetary material \citep[][]{Weber2022}. 
Further, it can be due to strong violation of the assumption of single scattering \citep[][]{Canovas2015,Pohl2017}, where the convolution of different polarisation states within the telescope's resolution cancel.
Finally, the simplest explanation for an intensity diminution is the lack of scattering material at the location of the decrement.
For AS~205S we treat the interpretation of those features with caution as some parts of the stellar polarisation may be unaccounted for by our reduction technique. Close to the star, this residual stellar polarisation may introduce artificial increments and decrements in the polarised light image.}

Besides the two discs, there is significant scattered light observed on large scales between and around the northern and southern objects. There is a spatially confined region that seems to connect AS~205N and AS~205S ($B$), to which we will refer as a ‘bridge'. This bridge is broader in the north and seems to taper towards AS~205S. The bridge shows well-defined edges, especially to the western direction. Additionally to the polarised intensity stemming from the bridge, we can distinguish a large-scale spiral in the diffuse region to the north of AS~205N ($LS$).

The overall structure observed in AS~205 is consistent with a hyperbolic stellar fly-by as dynamical origin \citep[see e.g. for $\beta=45^\circ$ shortly after pericentre, second row, central panel in Fig.~2 of][]{Cuello2019}. The clockwise direction of the spirals $S_1$ and $S_2$ around AS~205N suggests a counter-clockwise fly-by and that the periastron of the orbit (location of closest approach) has already been crossed. However, coupled binary formation from a common molecular cloud of high angular momentum cannot be ruled out either, as it can produce very similar features \citep[see Fig.~12 in][]{Bate2022}. We will further discuss the dynamical origin of AS~205 in Sec.~\ref{subsec:discussion_AS205}.

\subsubsection{Unresolved polarisation}
We place circular apertures over the adaptive optics residuals around AS~205N, rejecting image artefacts and areas that expose disc or bridge structures, to measure the unresolved polarisation attributed to the stars. We repeat the same for AS~205S. We find that AS~205N has an estimated $DoLP$ of $(0.21\pm0.04)\%$ with an $AoLP$ of $(59.2 \pm 5.6)^{\circ}$. The secondary shows a much higher $DoLP$ of $(0.96\pm0.08)\%$ and an $AoLP$ of $(26.0 \pm 2.4)^{\circ}$, where the errors are the standard deviations due to different values in measurements of each polarimetric cycle measured by IRDAP \citep[][]{vanHolstein2020}.

As the $DoLP$ and $AoLP$ of both stars are very different, we do not expect the unresolved polarisation to originate from interstellar dust between the source and the observer. Yet, the $DoLP$ of AS~205S does not exclude the possibility of dust grains being present with a projected distance too close to the stellar position in the image plane to be separately resolved. This could be due to disc material at unresolved scales close to the star. An inner disc would produce an $AoLP$ perpendicular to its $PA$. For AS~205S the angles are roughly consistent with close-in material aligned with the outer disc ($PA$=$109.6^{\circ}\pm1.8^{\circ}$). Yet, the $DoLP$ of AS~205S could also arise from dust grains in the outer disc area that are within the line-of-sight towards the star due to the disc's high inclination and vertical extent. 
Finally, considering the perturbed nature of the system, the unresolved polarisation may be partly due to sparse circumbinary material in the line-of-sight.

Independent of the components physical origin, the unresolved polarisation is present in the stellar halos and, thus, corrupts the polarised intensity that arises due to scattering in the regions of interest.
We describe the employed subtraction of unresolved polarisation in binary systems in Appendix~\ref{appendix:star_pol}.

\subsubsection{Interpreting the polarised intensity}
{For the interpretation of the $AoLP$, we consider two light sources in the system, AS~205N and AS~205S, even though AS~205S is a binary itself. The reason for this is that the upper limit for the angular separation between AS~205Sa and AS~205Sb \citep[$\sim 0\farcs015$,][]{Kurtovic2018} is much smaller than the relevant image scales and the light of the two southern stars is thus scattered under approximately the same angle. Yet, binary effects in scattered light intensity may still be relevant for self-shadowing \citep[][]{DOrazi2019} or shadowing by material close to the binary \citep[][]{Weber2022}.}

\begin{figure}
    \centering
    \includegraphics{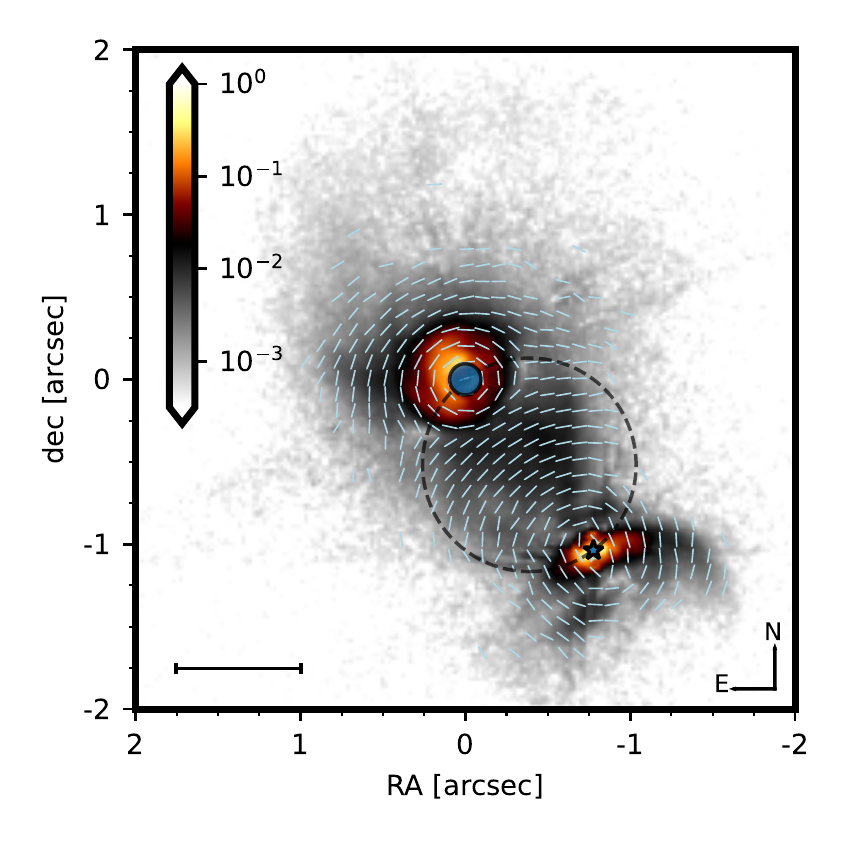}
    \vspace{-0.5cm}
    \caption{{Normalised polarised intensity of AS~205 in logarithmic scale, overplotted with AoLPs. Along the dashed circle the incident directions from AS~205N and AS~205S are perpendicular. The bar in the bottom left corner indicates the projected distance of 100$\,$au. The coronagraph area is masked by a circle, the location of AS~205S is indicated by a star.}}
    \label{fig:aolp_as205}
\end{figure}

If dominated by single-scattering events, the $AoLP$ is expected to be perpendicular to the direction of incoming light.
Fig.~\ref{fig:aolp_as205} shows that whereas the $AoLP$ in the discs is centrosymmetric towards the respective host star, it is centrosymmetric towards AS~205S in the bridging area. 
This exposes that the polarised light observed from the bridge is scattered light from the southern source. 
There are two possibilities why this could be the case: {\it (i)} the geometry of the system is such that the direct illumination is stronger from the south, or {\it (ii)} the scattering angle is favourable towards AS~205S such that the scattered light has a larger polarisation fraction \citep[e.g.][]{Min2012,Min2016,Stolker2016}.

Further, two intensity nulls appear to the west and south-east of the primary's location ($N_1$ and $N_2$ in Fig.~\ref{fig:mosaic}). A local lack of polarised intensity can be caused by several effects as discussed previously in Sec.~\ref{subsubsec:as205_observedstruc}.
Here, we interpret these features as an effect of the multiple light sources present in the system: the intensity vanishes where polarisation produced from the northern and southern source cancel via incoherent summation of their $Q$- and $U-$components.
The location of the intensity nulls is accurately consistent with our analytical expectation derived in Sec.~\ref{sec:two_light_sources}.
The quintessence of the analysis there is that the polarised intensity vanishes where {(\it i)} the incident directions towards both light sources are perpendicular and {(\it ii)} the stars' individual contribution to the polarised light are equal.
In Fig.~\ref{fig:aolp_as205} we overplot the polarised intensity image with a circle defined by the first condition. We can see that both intensity nulls fall accurately onto this circle.
The intensity nulls are located much closer to AS~205N than to AS~205S, which is consistent with the suggestion that the largest part of the inter-binary region is dominated by scattered light from the southern source which we previously derived from the alignment of the $AoLP$.
We will discuss this assessment further in Sec.~\ref{subsec:stellar_brightness}.

\subsection{SR~24}
The IRDIS detector has a fixed field-of-view of $11\farcs0 \times 11\farcs0$. The separation of the two main components in SR~24 is about $5\farcs1$ \citep[][]{GaiaDR3}. As the observation is centred on the northern source, the southern source is consequently very close to the detector's edge, which appears as a cut-off in the image shown in Fig.~\ref{fig:mosaic}. We, therefore, point out that the global structure is not entirely captured in the observation. The angular separation of SR~24N and SR~24S provides a lower limit of $511.4\,$au for their deprojected mutual distance.
Further, SR~24 is a very faint system in $H$-band, the signal-to-noise level is therefore about one order of magnitude lower than in the other two systems.

\subsubsection{Observed Structure}

Both components of SR~24 are surrounded by extended polarised scattered light coming from discs around SR~24N and SR~24S, respectively, as reported by \citet{Mayama2020}. As for the previous source, we detect a connecting bridge ($B$) for the first time in polarised light ({more clearly visible in an image of reduced Poisson noise in Appendix~\ref{appendix:SR24_0} and} previously observed in the NIR total intensity \citealp{Mayama2010}). 
For the disc around SR~24S, radial intensity profiles measured by \citet{Mayama2020} showed significant differences between the south-eastern and north-western direction, which the authors attributed to shadowing by an inner misaligned disc component.
In the SPHERE/IRDIS observation we can analyse the intensity profile with higher sensitivity. The IRDIS image shows the asymmetry addressed by \citet{Mayama2020}; a bright, smaller region to the south-east ($FS$) and a dim, elongated region to the north-west ($BS$). In contrast to the interpretation by \citet{Mayama2020}, here, we interpret the image as the observation of a very flared, inclined disc, of which we see the front side, $FS$, and partly the back side, $BS$ (more on this in Sec.~\ref{subsubsec:discussion_SR24}).

The northern disc constitutes a circumbinary disc around the components SR~24Na and SR~24Nb. The individual stars are not visible in the image as they are hidden behind the coronagraph. 
Also the northern disc displays strong asymmetrical structure with the presence of extended scattering north-west of the stars ($S_1$), tracing a spiral arm that opens in counter-clockwise direction and is opposed by a spiral arm to the south-west of SR~24N ($S_2$).
The southern spiral arm smoothly merges into the bridge towards SR~24S.

\subsubsection{Unresolved polarisation}\label{subsec:SR24_unresolved_pol}
The extraction of the stellar polarisation from its halo was not feasible in this case, as there is no region in the image where we could confidently isolate the halo from local dust scattering for neither of both stars. We chose instead a different technique to estimate the polarisations: We centred a small mask on the respective star and measured $Q$ and $U$ in each pixel. We assume that those two components are dominated by the unresolved polarisation within this mask. We average over all attributed pixels and take the standard deviation as the error of the $Q$ and $U$ measurement. We thus estimate a $DoLP$ of $(1.24\pm0.16)\%$ and an $AoLP$ of $(8\pm5)^\circ$for SR~24N and a $DoLP$ of $(1.1\pm1.0)\%$ and $AoLP$ of $(3\pm22)^\circ$ for SR~24S.
The low level of confidence for the values of the southern source motivates us to include the uncorrected image in Appendix~\ref{appendix:SR24_0}.
When looking at Fig.~\ref{fig:aolp_sr24}, we see that the $AoLP$s in image areas close to the southern star are centrosymmetric towards SR~24S and correspongingly, the $AoLP$s around the northern binary are centrosymmetric towards SR~24N. The transition seems to occur within the bridging region. 
We note, that we do not observe strong deviation from symmetry around SR~24S as described in \citet[][]{Mayama2020} who ascribed this to illumination from the northern component. \begin{figure}
    \centering
    \includegraphics{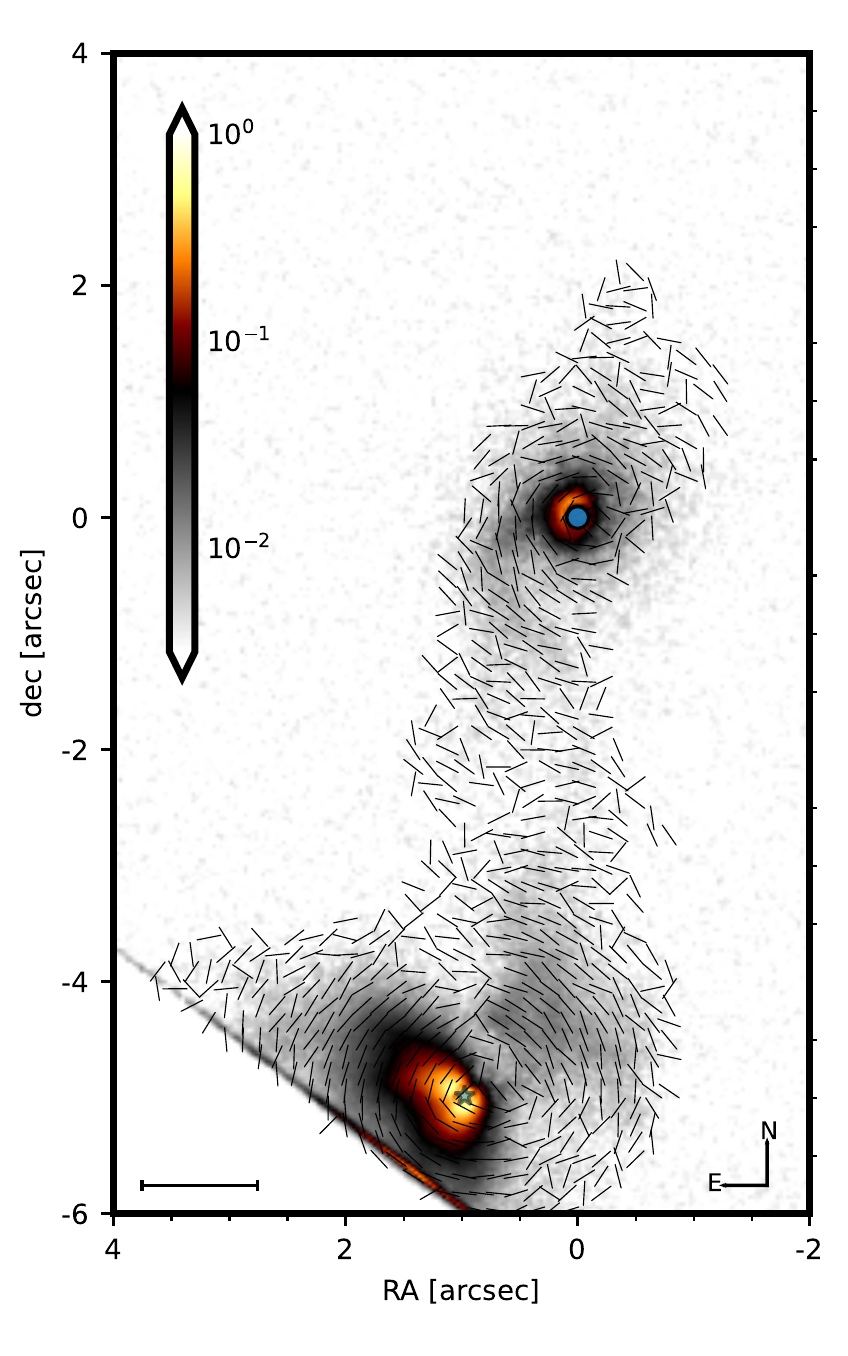}
    \vspace{-0.8cm}
    \caption{{Normalised polarised intensity of SR~24 overplotted with $AoLP$s shown in logarithmic scale. The bar in the bottom left corner indicates the projected distance of 100 au. The coronagraph area is masked by a circle, the location of SR~24S is indicated by a star.}}
    \label{fig:aolp_sr24}
\end{figure}

\subsubsection{ALMA continuum structure}
\begin{figure}
    \centering
    \includegraphics{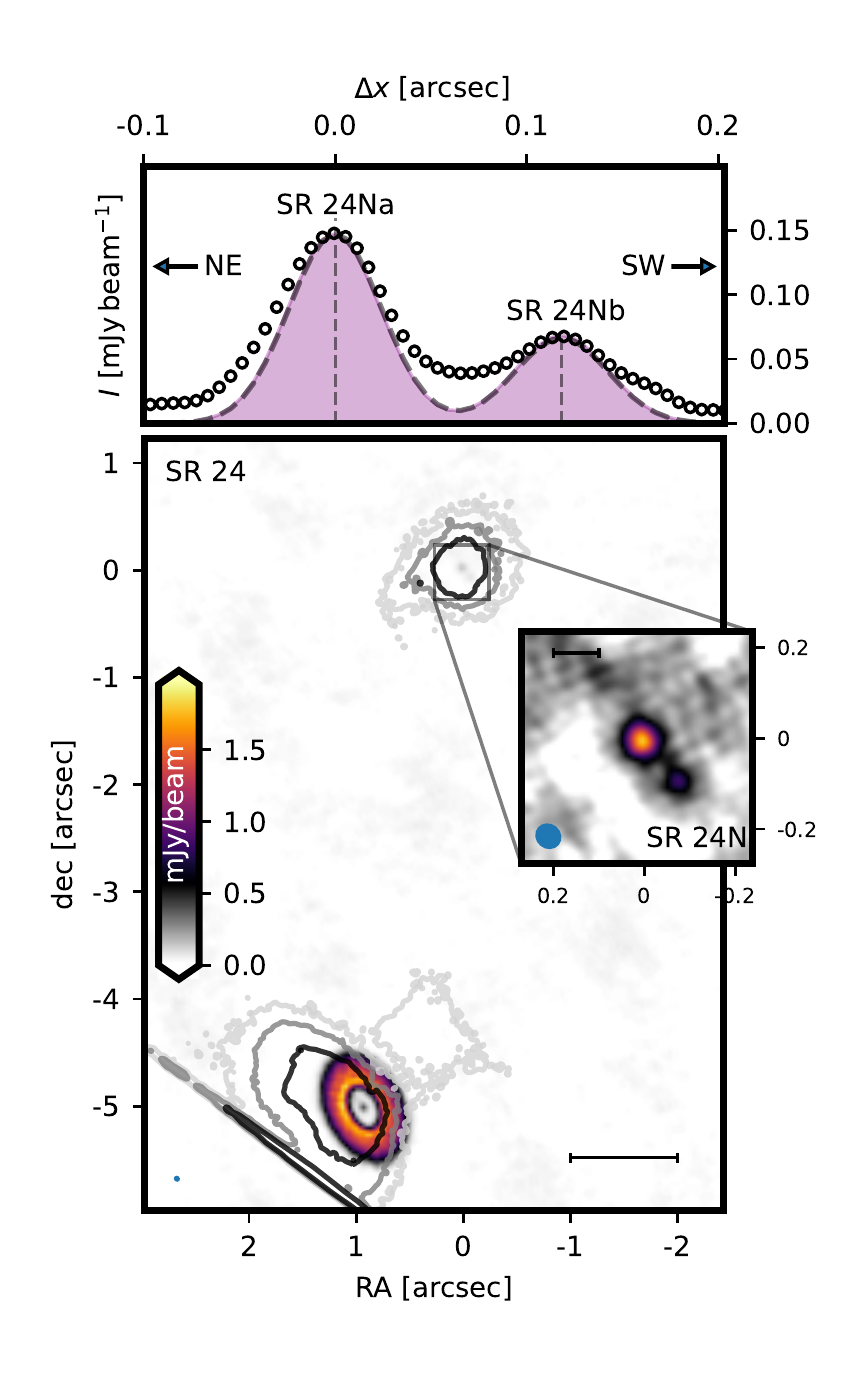}
    \vspace{-0.5cm}
    \caption{Lower panel: ALMA band 6 intensity map for SR~24. Contours trace the $H$-Band polarised intensity presented in Fig.~\ref{fig:mosaic} for comparison at levels of 0.25\%, 1\% and 5\% of the peak intensity. The bar in the bottom right corner shows the projected distance of 100$\,$au. The inset zooms in on the location of SR~24N. Contrast in the inset is increased by a factor of ten, the two sources SR~24Na and SR~24Nb are detected with $\sim12\,\sigma_{\rm rms}$ and $\sim 5\,\sigma_{\rm rms}$, respectively. The bar in the top left corner of the inset corresponds to 10$\,$au. Top panel: Intensity along the connecting line between SR~24Na and SR~24Nb, north-east is to the left, south-west to the right. The circles show the ALMA band 6 data points, the dashed lines show the beam size of the observation.}
    \label{fig:cont_sr24}
\end{figure}

The re-reduced ALMA band~6 data of SR~24 \citep[previously published in][]{Cieza2021} are shown in Fig.~\ref{fig:cont_sr24}. The main panel shows the continuum map centered on SR~24S (see Sec.~\ref{subsec:ALMAdata} for details), overplotted with contours of polarised light intensity for visual guidance. Analysis of continuum emission around SR~24S was already performed on other data sets \citep[][]{vanderMarel2015,Pinilla2017,fernandez-Lopez2017,Pinilla2019}. An unresolved millimeter emission around the northern component SR~24N was detected with ALMA at 1.3$\,$mm by \citet{fernandez-Lopez2017}.
Our re-analysis of the ALMA continuum data from the ODISEA programme resolves this emission for the first time into two individual, circumstellar components from SR~24Na and SR~24Nb (see inset and top panel of Fig.~\ref{fig:cont_sr24}). The contrast in the inset of Fig.~\ref{fig:cont_sr24} is increased by one order of magnitude to highlight the two components in SR~24N. 
{The top panel of Fig.~\ref{fig:cont_sr24} shows the intensity along the connecting line between the two sources, SR~24Na and SR~24Nb, and compares the profile of their emission with the beam size of the observation (colour-filled area). The peak emissions of the two discs are separated by $0\farcs116$.}
{The estimated root mean square (rms) noise of the observation is $\sigma_{\rm rms}=12\,$µJy$\,$beam$^{-1}$. Both detections are significant, with peak fluxes of $F_{\rm peak,Na}\sim 12\,\sigma_{\rm rms}$ and $F_{\rm peak,Nb}\sim 5\,\sigma_{\rm rms}$.}
We will compare the measured positions from this ALMA image with other astrometrical measurements in Sec.~\ref{subsubsec:discussion_SR24}.\\
{Using the CASA task {\tt imfit}, we fit a two-dimensional elliptical Gaussian in the image plane to the disc component around SR~24Na that yields a size of $0\farcs084\times 0\farcs075 \pm 0\farcs005$. Thus, the fitting suggests that the disc component around SR~24Na is resolved. We deconvolve the size from the synthesised beam ($0\farcs060\times0\farcs055$) to report the full width at half maximum (FWHM) of the disc of FWHM$_{\rm Na}=(45\pm5)\,$mas.} 

{The fainter emission from SR~24Nb makes its modelling more challenging: the lower signal-to-noise and the vicinity of the brighter companion SR~24Na likely lead the Gaussian fit to overestimate the size of the source (see Fig.~\ref{fig:cont_sr24}, upper panel). In these circumstances, we can consider the resulting deconvolved FWHM$_{\rm Nb} = (69\pm9)\,$mas to be only an upper limit for SR~24Nb.}

{We estimate the dust masses present in the circumstellar discs for two different effective temperatures and list them in Table~\ref{tab:SR24Ndiscs}. The dust mass in each discs is of the order of one lunar mass, ${\rm M}_{\rm Moon}$, and therefore about three orders of magnitude inferior with respect to the estimated dust mass around SR~24S \citep[][]{Cieza2021}. The masses were calculated using the equation:}
\begin{equation}\label{equ:dust_mass}
    M_{\rm dust} = \frac{F_{\nu}d^2}{\kappa_{\nu}B_{\nu}(T_{\rm dust})}\,,
\end{equation}
{where $F_{\rm tot}$ is the integrated disc flux at the observing frequency, $d$ is the distance and $B_\nu(T_{\rm dust})$ the Planck function at the given dust temperature and observing frequency. Note, that to calculate $F_{\rm tot}$ for SR~24Nb, we assume the source to be unresolved.}
To compare to other ODISEA sources, we assume the same dust opacity of $\kappa_\nu = 2.3\,{\rm cm}^2\,{\rm g}^{-1}$ and, in the first estimate, a temperature of $T_{\rm dust}=20\,$K \citep[][]{Cieza2021}. As the emission traced here is very close to the respective stellar host, we present an additional mass estimate considering a higher temperature for the Planck function of $T_{\rm dust}=100\,$K.
\begin{table}
\begin{center}
\caption{{Circumstellar continuum emission around SR~24N.}}
\begin{tabular}{ |c||c|c|c|c|c| }\label{tab:SR24Ndiscs}
  & FWHM  & $F_{\rm peak}$ & $F_{\rm tot}$  & $M_{{\rm d},20\,{\rm K}}$& $M_{{\rm d},100\,{\rm K}}$\\
  & [mas] & [µJy$\,{\rm beam}^{-1}$] & [µJy]  & [${\rm M}_{\rm Moon}$]& [${\rm M}_{\rm Moon}$]\\
 \hline
 \hline
  {\it Na} & $45\pm5$ & $150\pm4$ & $229\pm9$ & $5.3\pm0.4$ & $0.84\pm0.06$ \\
 {\it Nb} & $<69\pm9$& $64\pm4$& $64\pm 4$ & $1.5\pm0.1$& $0.24\pm0.02$ \\
 \hline
 \hline
\end{tabular}
\end{center}
\end{table}
We want to highlight that the systematic uncertainty of the disc masses is expected to be much higher due to its dependency on opacities and temperature, i.e. on the assumed dust and disc model, as can be appreciated for example in \citet{Guidi2022}.

\subsection{FU~Orionis}
The image of scattered light in the FU~Orionis system (bottom left panel of Fig.~\ref{fig:mosaic}) shows a highly sub-structured and perturbed pattern. This SPHERE/IRDIS $H$-band image confirms the structures observed with Gemini/GPI in $J$-band \citep[see Fig.~6 in][we index the observed features with the same letters]{Laws2020}: the small companion disc ($A$), the bright arm to the north-east ($B$), several dark lanes ($C$) super-imposed on a tail of scattered light ($D$) in the north-west and west of the primary, the diffuse scattering to the south-west ($E$). 

The SPHERE/IRDIS observation was taken about one year before the Gemini/GPI observation (December 2016 and January 2018, respectively). The agreement of bright features and dark lanes between those two images indicates that they are not very variable on such small time-scales, which was previously speculated due to differences with HiCIAO observations presented in \citet{Takami2018} (e.g. scattered light excess in the western region). The differences between the GPI and HiCIAO images may be caused by uncorrected instrumental effects in HiCIAO \citep[as mentioned in][]{Laws2020}.

\begin{figure}
    \centering
    \includegraphics{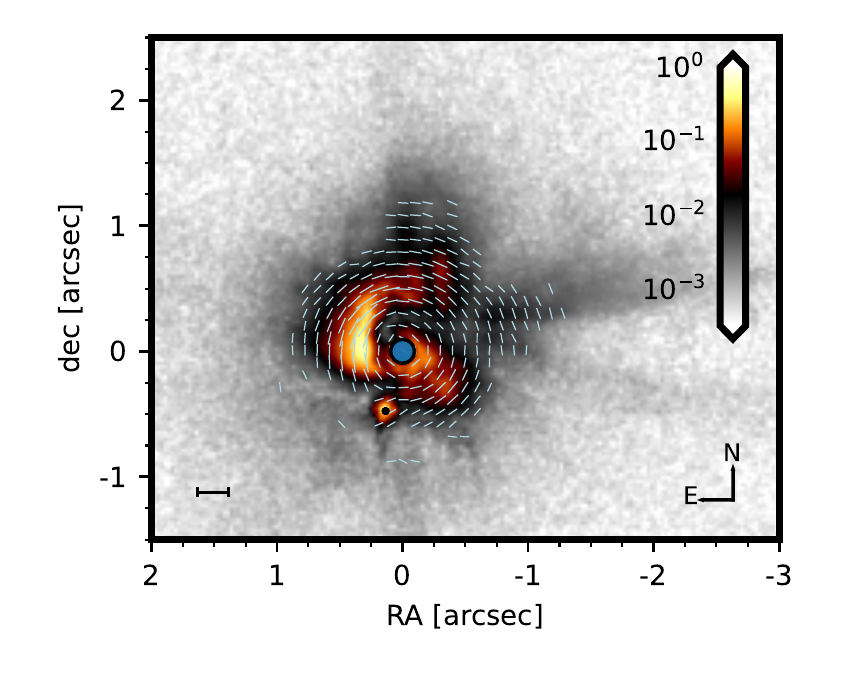}
    \vspace{-0.5cm}
    \caption{{Normalised polarised intensity of FU~Orionis overplotted with AoLPs shown in logarithmic scale. The bar in the bottom left corner indicates the projected distance of 100 au. The coronagraph area is masked by a circle and centred on FU~OriN, the location of FU~OriS is indicated by a star.}}
    \label{fig:aolp_fuori}
\end{figure}

The mostly centrosymmetric $AoLP$ pattern overplotted in Fig.~\ref{fig:aolp_fuori} reveals that most of the polarised light is scattered from the northern primary. An exception form the image regions very close to the secondary; its disc (as already observed with Subaru/HiCIAO in \citealp{Liu2016} and \citealp{Takami2018}) and to some smaller degree the southern parts of the bright arm \citep[consistent with significant $U_\phi$ measurements from this region,][see also Appendix~\ref{appendix:QUphi}]{Laws2020} and the eastern parts of the diffuse region in the south-west. 
This means that large parts of the polarised intensity in the image are entirely due to scattering from FU~OriN. It is, therefore, likely that the dark lanes north-west of the primary disc ($C$) are not caused by polarisation cancellation but are rather due to a complicated geometrical arrangement of the dust grains as discussed in \citet{Laws2020}.

\subsubsection{Unresolved Polarisation}
In FU~Orionis the southern source lies within or very close to the structure around the northern star, so that the measurement of the stellar polarisation is again not possible with standard procedures. Instead we measure again the $Q$ and $U$ values in a very confined mask centred on FU~OriS, similar to the procedure adopted for SR~24. 
We measure stellar $DoLP$s of ($0.27\pm0.06$)$\%$ and ($1.24\pm0.11$)$\%$ for the northern and southern component, respectively, with $AoLP$s of ($161\pm7$)$^\circ$ and ($112\pm3$)$^\circ$. The values for FU~OriS should be treated with care as even after applying the putative correction for unresolved polarisation the $AoLP$ pattern is not centro-symmetric towards the position of the star, suggesting that there are strong contributions from the northern light source, secondary scattering or foreground material that have not been capture by the applied reduction procedure.

\section{Discussion}\label{sec:discussion}
\subsection{Stellar brightness}\label{subsec:stellar_brightness}
The observations presented in Sec.~\ref{sec:results} inform about dust structures in three twin disc systems. We highlighted that the angle and intensity of linearly polarised light between two light sources are strongly dependent on the morphology of the system. There are several things to be taken into account when analysing the influence of the two different light sources. One important quantity is the brightness ratio of the two stars in the relevant observational band. 
The NACO measurements presented in Table~\ref{tab:mags} show that AS~205N is one magnitude brighter in $H$-band than AS~205S. 
{In Sec.~\ref{sec:results}, however, we saw that the $AoLPs$ in the region between the stars indicate prevalence of light from the southern star, consistent with the proximity of the two $PI$ nulls ($N_1$ and $N_2$) to AS~205N if interpreted as polarisation cancellation.} As the brightness measurement in $H$-band shows the dominant scattering from the southern source in the bridging region is not due to an inherent brightness difference between the two sources, on the contrary. This leaves three possibilities (or a combination thereof) of why scattering in the bridge area is dominated by the fainter southern source, all of which point to a geometrical effect of the objects positioning: {\it(a)} the northern source does not directly illuminate this material. This could be the case if the bridging area is in the foreground with respect to the northern star, or if it is in the shadow of the disc around AS~205N. {\it (b)} the deprojected distance of the bridge is much closer to the southern source, {\it (c)} the scattering angle is favourable towards the southern star and promotes the polarisation from its incident direction. In contrast to case {\it (a)}, here, the northern source could still illuminate the bridge, only the scattering of its light is not efficiently polarised.

In SR~24 the $H$-band brightness contrast between the two light sources is similar as in AS~205; here the southern source is brighter by about one magnitude. 
{The $AoLP$ presented in Fig.~\ref{fig:aolp_sr24} does not reveal from which star the light is predominantly scattered. The reason for this is that the signal-to-noise level of the image is much lower than for AS~205 and that the bridge is limited to a narrow region along the connecting line between SR~24N and SR~24S, such that the scattering of light from both stars produces similar $AoLP$s in this region.}

For FU~Orionis the brightness contrast is even more pronounced. As mentioned in Sec.~\ref{sec:targets}, the observed light from FU~OriS is supposedly subjected to strong extinction along the line-of-sight leading to the controversy that even as the star is estimated to be the more massive component, it appears dimmer by about four orders of magnitude. \citet{Beck2012} comment that the flux attenuation measured in different NIR bands are prominently incompatible with obscuring material that follows an ISM extinction law. One possibility might be that FU~OriS is veiled by disc material around FU~OriN, which would constrain their relative positioning in three-dimensional space. It would suggest that FU~OriS is in the background with respect to the bright scattering material that is centred around FU~OriN.
\subsection{Comparison to other data}
In the following we compare the presented scattered light images to archival data.
\subsubsection{AS~205}\label{subsec:discussion_AS205}
As part of the DSHARP programme \citep[][]{Andrews2018} AS~205 was observed in high-resolution with ALMA band 6 \citep{Kurtovic2018}.
\begin{figure*}
    \centering
    \includegraphics{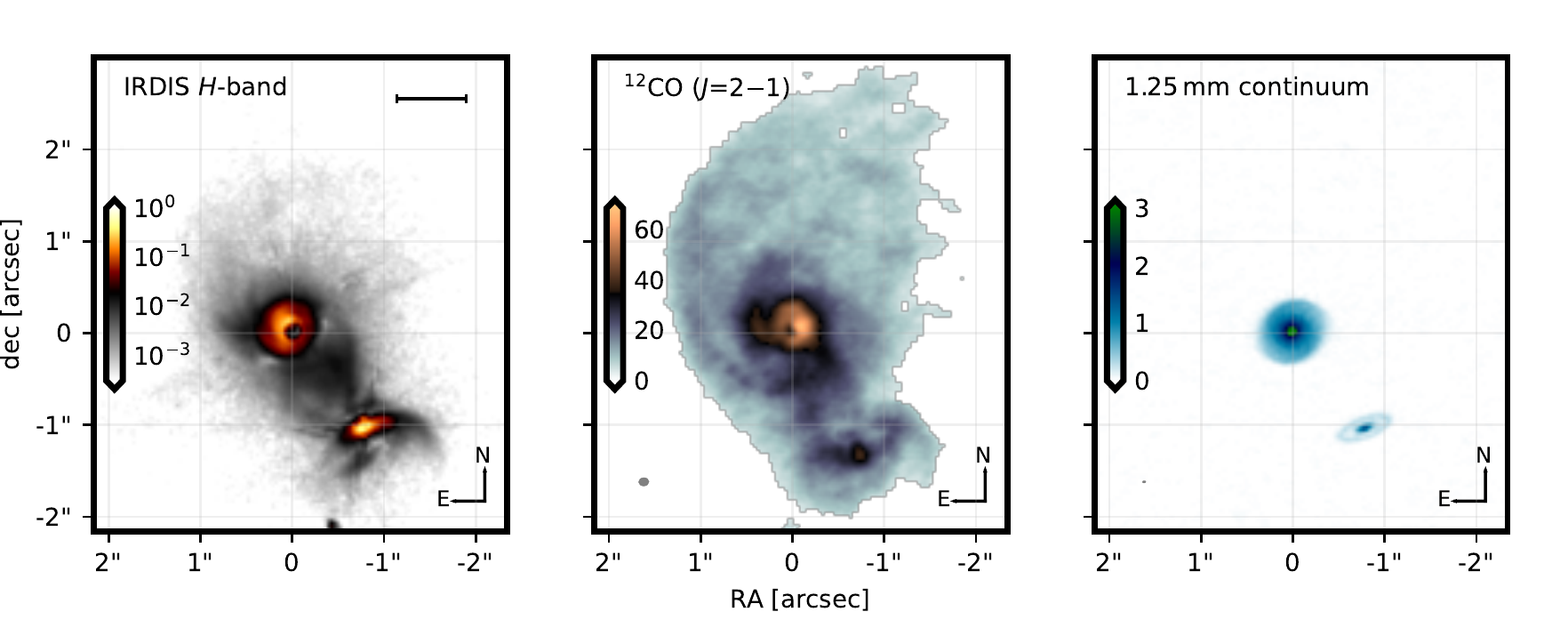}
    \vspace{-0.3cm}
    \caption{{Comparing different observations of AS~205: the left panel shows the NIR polarised light intensity, observed with SPHERE/IRDIS in $H$-band ($\lambda_{\rm obs}=1.625\,$µm). The intensity has been normalised with its maximum value, and we show the image in logarithmic stretch, indicated by the colour bar. The bar in the top right corner corresponds to a projected distance of 100$\,$au. The central panel shows the ALMA $^{12}$CO~($J$=2--1) moment 0 image and the right panel shows the ALMA band~6 ($\lambda_{\rm obs}=1.25\,$mm) continuum, For both ALMA images the colours are in linear stretch and in units of mJy$\,$beam$^{-1}$. The beam size is shown in grey in the bottom left corner of the images. The ALMA data were originally published in \citet{Kurtovic2018} (programme ID 2016.1.00484.L). The data were taken at different epochs (19$^{\rm th}$ of May, 2017 for IRDIS and 29$^{\rm th}$ of September, 2017 for ALMA).}}
    \label{fig:P_CO_cont_AS205}
\end{figure*}
In Fig.~\ref{fig:P_CO_cont_AS205} we show the polarised light image next to a $^{12}$CO moment 0 map and the band 6 continuum observation. All images are to scale with grid lines facilitating direct comparison. The image illustrates that the global structure of polarised scattered light traces well the observed $^{12}$CO emission. The CO moment 0 map also shows the large scale spiral to the north ($LS$) and the bridging material ($B$). In the south, however, the CO emission is off-centred from the star. Locally, around the stars, the IRDIS image shows stronger disc features. The ALMA band 6 continuum emission is restricted to the individual discs, without any visible inter-binary material. This is expected, as evolved particles' radial inward drift is promoted in multiple systems due to the tidal interaction \citep[][]{Zagaria2021a}. 

\begin{figure}
    \centering
    \includegraphics{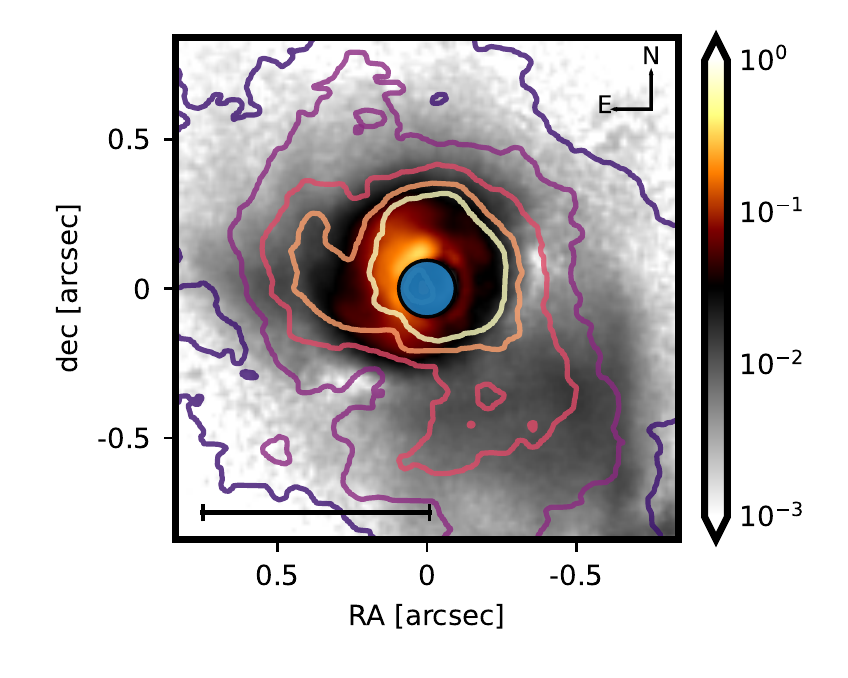}
    \vspace{-0.5cm}
    \caption{{Zoom-in on AS~205N. The colour map shows the normalised polarised intensity in logarithmic stretch, the contours show the $^{12}$CO($J$=2--1) moment 0 at levels of 20\%, 30\%, 40\%, 50\% and 60\% of the peak value. The central circle shows the coronagraph used for the DPI observation.}}
    \label{fig:P_CO_comp_AS205}
\end{figure}
{In Fig.~\ref{fig:P_CO_comp_AS205} we show a magnified direct comparison of the polarised intensity and CO moment 0 around AS~205N. This comparison shows that the prominent spiral arm to the north-east visible in polarised scattered light ($S1$) is also present in the gas structure. Further, we can see that the polarisation null located to the south-east ($N_1$) coincides with a local indentation of gas emission. This questions the interpretation of the lack of polarisation as being due to polarisation cancelling. Line emission is not affected by this feature. If the lack of intensity in the polarised scattered light and gas emission share the same origin, a local decrease of density or temperature would be the more plausible explanation. The other polarised intensity null on the western side of the disc ($N_2$) does not coincide with any local decrease in the gas emission.}
\paragraph*{Gravitationally bound binary or Fly-by?}
We attempt to constrain the nature of AS~205 (bound binary or hyperbolic fly-by) by comparing the sources' relative velocity to the escape velocity of the system. The escape velocity can be calculated as:
\begin{equation}
    v_{\rm e}=\sqrt{\frac{ 2GM_{\text{total}} }{ R }}
\end{equation}
\noindent where $G$ is the gravitational constant, $M_{\text{total}}$ is the sum of the individual masses, and $R$ is the physical distance between the sources. Whereas the individual masses can be estimated from observations, the physical distance depends on the unconstrained line-of-sight distance. The projected distance, on the other hand, has been measured ($1\farcs34$ corresponding to $177\,$au at $132\,$pc from \citealp{GaiaDR3}), which constitutes the lowest possible value for $R$.

\begin{figure}
    \centering
    \includegraphics{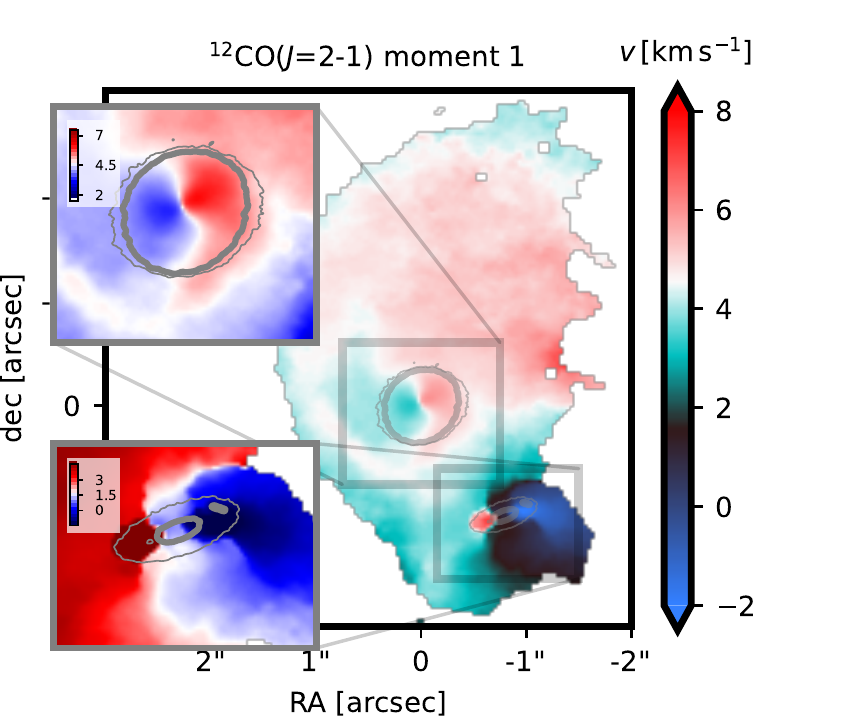}
    \caption{Showing the global $^{12}$CO($J$=2--1) moment 1 in AS~205. The white colour is centred on the motion of AS~205N, black is centred on AS~205S. The insets show zoom-ins onto the two discs, highlighting their rotation in the classical (blue-whit-red), adapted colour scale. The grey contours in the insets trace the band 6 continuum emission at 5$\sigma_{\rm rms}$ (thin line) and 20$\sigma_{\rm rms}$ (bold line)}.
    \label{fig:AS_kine}
\end{figure}

Fig.~\ref{fig:AS_kine} shows a map of the $^{12}$CO(J=2-1) kinematics for AS~205.
We analyse these data to estimate the dynamical stellar mass and central velocity of each disc making use of the \texttt{eddy}-package \citep{eddy}. This method, however, has many caveats when applied to the AS\,205 system, due to the highly perturbed surface emission with non-circular Keplerian motion (as shown in the insets of Fig.~\ref{fig:AS_kine}). To minimise the number of free parameters, we do not fit for the disc geometry, position or surface height. Instead, we fix the centre and position angle of the disc from the dust continuum emission.   
{\citet{Kurtovic2018} computed the discs' inclinations from different methods: by fitting an elliptical Gaussian to the continuum emission, or by fitting the geometries of continuum disc features, such as the spirals for AS~205N (assuming a logarithmic model or an Archimedean model) and the ring for AS~205S \citep[see][for details of each method]{Kurtovic2018}. 
Depending on the method, their results for the disc inclination differ by a few degrees. Thus, we calculate the dynamical stellar mass for several values of the disc inclination, sampled around the measurements in \citet{Kurtovic2018}.}

Further, the inferred dynamical mass is also affected by the chosen size of the mask around the stars where the kinematics are modelled. 
\begin{figure*}
    \centering
    \includegraphics[width=\textwidth]{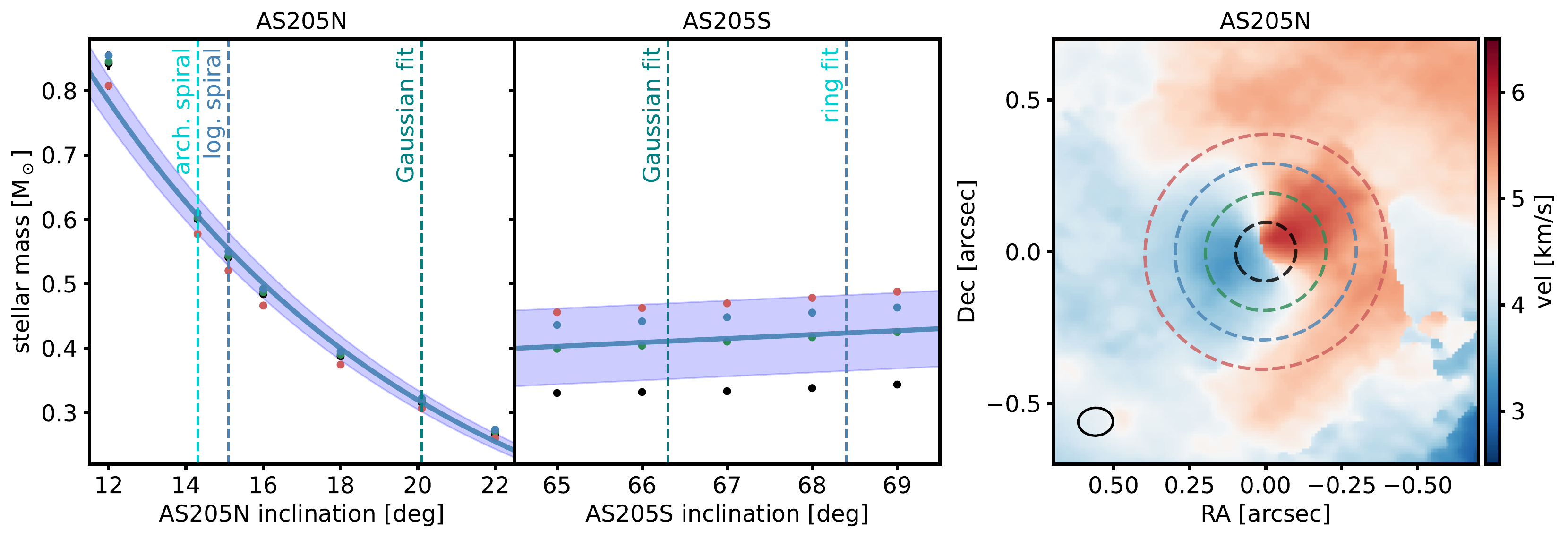}
    \caption{{The left and central panel show dynamical stellar mass estimates for AS~205N and AS~205S, respectively, measured with the {\tt eddy}-package \citep[][]{eddy}. The stellar masses were calculated for different assumed disc inclinations, spread around the estimates provided in \citet{Kurtovic2018} from different methods, for which the measured inclination values are indicated by vertical dashed lines. At each assumed inclination we varied the size of the mask to which the dynamical measurement was limited between radii of $0\farcs1$, $0\farcs2$, $0\farcs3$ and $0\farcs4$. AS a visual example, the right panel shows the moment 1 map centred on AS~205N with the different mask sizes indicated by dashed circles. The colour of the data points in the two left panels corresponds to the colour of the applied mask in the right panel.}}
    \label{fig:AS_kine_masks}
\end{figure*}
This can be better appreciated from Fig.~\ref{fig:AS_kine_masks}, where in the right panel the line of nodes (traced by white colour) is more or less perturbed as a function of the distance to the disc centre. For each disc, we fit models using mask radii of [$0\farcs1$, $0\farcs2$, $0\farcs3$, $0\farcs4$], shown for the example of AS~205N in the right panel of Fig.~\ref{fig:AS_kine_masks} as dashed circles. The mask size variation is combined with the grid of inclinations that was explored for each disc. In total, we explored 28 different fits for AS~205N and 24 for AS~205S. The left and central panel of Fig.~\ref{fig:AS_kine_masks} show how the stellar mass measurement changes with mask size and assumed disc inclination, for AS~205N and AS~205S, respectivly.

For AS\,205N, we find the central mass to be weakly dependent on the mask size, but strongly dependent on the disc inclination. Depending on the method to infer the inclination, \citet{Kurtovic2018} provide three different inclination values: $(20.1\pm3)^\circ$ from an elliptical Gaussian fit to the dust continuum, $15^\circ.1^{+1.9}_{-3.2}$ from fitting a logarithmic spiral, and $14^\circ.3^{+1.3}_{-5.3}$ from a Archimedean spiral to the dust continuum. The inclination values obtained from the different spiral models are consistent with each other. We take the average of the masses recovered with the spirals' inclinations as the mass for AS\,205. It has to be kept in mind, however, that this mass was recovered under the assumption that the dust spirals follow one of those models. Then, the estimated mass from kinematics is $M_{N\star}=(0.58\pm0.05)\,{\rm M}_\odot$, with a velocity in the line of sight of $v_{N,los}=(4.49\pm0.02)\,{\rm km}\,{\rm s}^{-1}$. The left panel of Fig.~\ref{fig:AS_kine_masks} shows that the Gaussian inclination measurement would lead to a considerably lower stellar mass.

The mass estimate for AS~205S depends only weakly on the assumed inclination of the disc, but shows a strong dependence on the radius of the mask. It is possible that this is because of the high disc inclination, and the distribution of the perturbed material around it. To estimate the line-of-sight velocity, we decided to consider the values obtained with the smallest mask ($0\farcs1$), as those are the models that trace the emission of the gas closer to the central spectroscopic binary. We constrain a combined mass of $M_{S\star\star}=(0.42\pm0.06)\,{\rm M}_\odot$ for the southern binary, with a velocity along the line-of-sight of $v_{S,los}=({1.75\pm0.01})\,{\rm km}\,{\rm s}^{-1}$. We want to emphasise that the given errors of the stellar mass measurements arise from the pixel sensitivity, but do not comprise any systematic errors which we expect to dominate the uncertainty (such as mask size and inclination as previously discussed in this section). We further stress that the {\tt eddy}-package, used to constrain the masses, assumes pure Keplerian motion from which the perturbed velocity profiles in the two discs (seen in the inset of Fig.~\ref{fig:AS_kine}) deviate. A complex morphology of the emission produces a systematic error of this mass measurement. The model from the {\tt eddy}-package is just a flat disc, and so the uncertainties are underestimated.

The total mass estimate from CO kinematics for the AS~205 is $M_{\rm tot}=(1.0\pm0.11)\,{\rm M}_\odot$. This value is significantly smaller than the value given in literature of $M_{\rm tot} = 2.15\,{\rm M}_\odot$, which was obtained by fitting spectroscopic data to a stellar evolution model \citep{Eisner2005,Andrews2018}.

Besides the on-sky positions of the stars, \citet{GaiaDR3} provides values for proper motion in the image plane for AS~205N and AS~205S. These values are currently not very reliable, as indicated by the large {\tt RUWE} values (4.0 for AS~205N and 1.7 for AS~205S).
The uncertainties in the following analysis do not take this systematic error into account, therefore, the results should be taken with caution and reliable conclusions have to wait for future astrometric improvement.

The relative motions of AS~205S with respect to AS~205N are listed as $v_{\rm RA} = (-2.63\pm 0.28)\,{\rm mas}\,{\rm yr}^{-1}$ and $v_{\rm dec} = (3.79\pm 0.21)\,{\rm mas}\,{\rm yr}^{-1}$. This gives an absolute value of proper motion of $(4.61\pm0.33)\,{\rm mas}\,{\rm yr}^{-1}$ in the image plane, corresponding to $(2.90\pm 0.21)\,{\rm km}\,{\rm s}^{-1}$. Together with the line-of-sight velocity inferred from gas kinematics we can hence determine the full proper motion of both binary components, most importantly finding an absolute relative velocity of ($5.64\pm0.24$)\,${\rm km}\,{\rm s}^{-1}$.
Then, the biggest uncertainty factors in the characterisation of the scenario of stellar interaction (bound binary or hyperbolic fly-by) are the systems' total mass and the line-of-sight distance between its binary components, both needed to establish the nature of the discs' interaction. 
\begin{figure}
    \centering
    \includegraphics[width=\columnwidth]{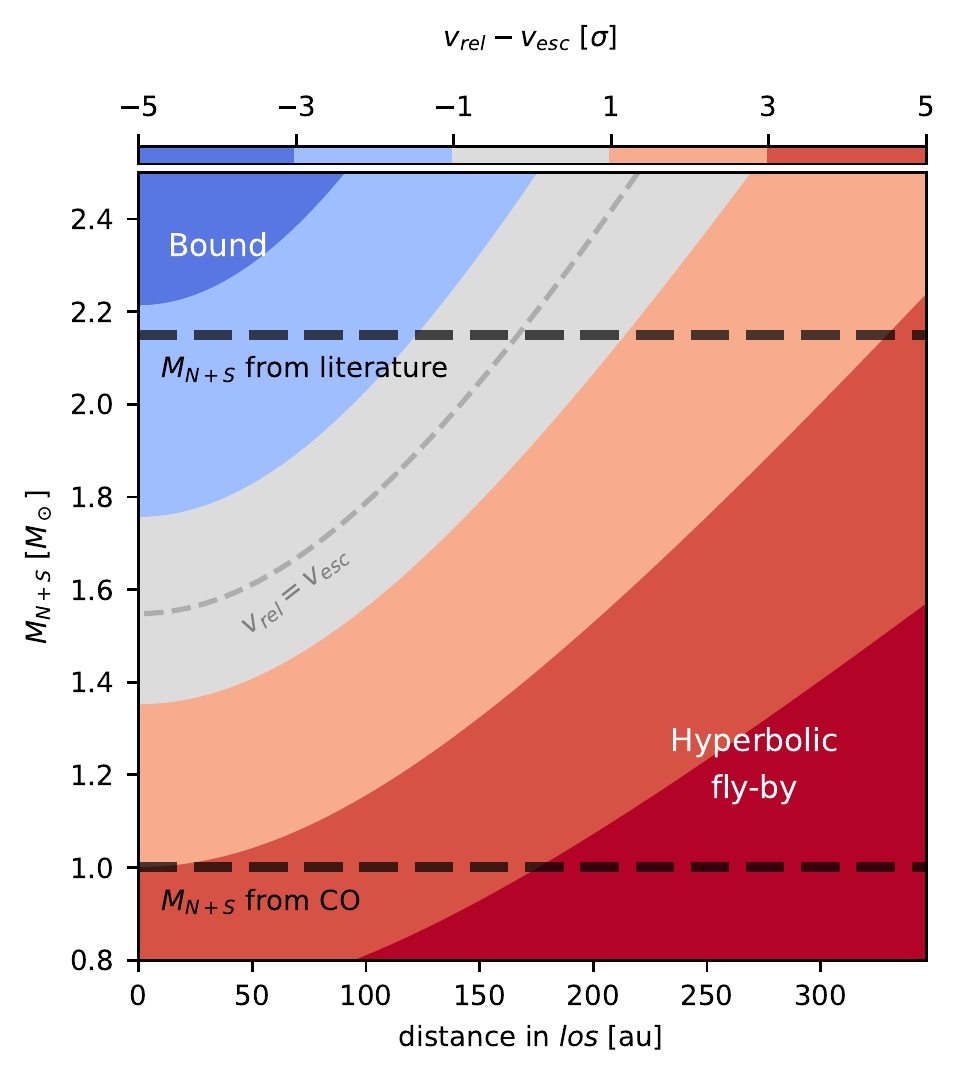}
    \vspace{-0.5cm}
    \caption{Comparison of relative velocity between AS~205N and AS~205S to the system's escape velocity, depending on its total mass and the components' separation along the line-of-sight ($los$). The velocity difference is given as a multiple of the uncertainty of the relative velocity inherited from Gaia measurements of proper motion \citep{GaiaDR3}. The grey dashed line shows where the velocities are equal; parameter combinations to the top-left are for a bound system, to the bottom-right are unbound and indicate, therefore, a hyperbolic fly-by. The upper horizontal dashed line marks the mass estimate found from spectroscopy \citep{Eisner2005,Andrews2018}, the lower dashed line marks our estimate from CO kinematics.}
    \label{fig:bound_unbound}
\end{figure}
In Fig.~\ref{fig:bound_unbound} we show how the dynamical state of the system can be classified, depending on those two variables.

Our constrained total mass differs from the mass found in literature (2.15$\,M_\odot$) by more than a factor of two. But also the literature mass inferred from spectroscopic data is not sufficiently reliable as it depends on stellar evolution models that introduce uncertainties to the calculation.
Stars undergoing an events of temporarily increased accretion, as suggested to occur episodically for young eccentric binaries \citep[e.g.][]{Kuruwita2020} or during a stellar fly-by \citep[e.g.][]{Borchert2022}, is one possible reason for a categorical overestimation of stellar masses in multiple systems when comparing to a stellar evolution track \citep{Baraffe2009,Jensen2018}.
For the dynamical mass estimate, future observations of tracers closer to the discs' mid-plane, such as the CO isotolopogues $^{13}$CO and C$^{18}$O could add better constraints as their molecular line emission stems from material  that is expected to be less affected by the stellar perturbation.
For AS~205N the disc inclination needs to be constrained within an error of $<1^\circ$ in order to recover a meaningful dynamical mass, a future challenge that should also consider a possible warp in the mid-plane geometry.

Nevertheless, our current mass estimate suggests that the kinetic energy of the system is larger than its potential energy, i.e. that the stars are not gravitationally bound.
This directly promotes the fly-by scenario, independent of the line-of-sight distance. Fig.~\ref{fig:bound_unbound} shows that the system is expected to be unbound even for a total mass of $1.5\,M_\odot$.  

The relative proper motion measured by Gaia suggests that AS~205S is moving towards the north-west with respect to AS~205N, consistent with a counter-clockwise fly-by \citep{Cuello2022}.
Also the spiral arms opening in clockwise direction around AS~205N are in agreement with this scenario. For an indubitable discrimination between the bound-orbit and fly-by scenarios, more reliable proper motion and mass measurements are needed.

\subsubsection{SR~24}\label{subsubsec:discussion_SR24}
Both components of SR~24 have been observed in polarised light with the Subaru/HiCIAO instrument \citep[][]{Mayama2020}. We reached a higher signal-to-noise level in our observations which allows us to additionally detect polarised signal from the bridging structure ($B$) between the northern binary and the southern star and additional components around the individual light sources. In comparison to the HST optical image, the measurement of $Q$ and $U$ components enable us to measure the $AoLP$ in the discs and in the bridge. The polarised light image further facilitates the analysis of the disc structure closer to the location of the stars where the HST image is dominated by the stellar halo. This is especially interesting with respect to the southern source, that appears extremely flared and separated into visible front and back sides.

\citet{Schaefer2018} suggested that unresolved ALMA continuum emission around SR~24N should be linked to a circumstellar disc around at least one of its components.
We separated the signal around SR~24N for the first time to reveal two small circumstellar discs around SR~24Na and SR~24Nb. We do not  detect any relevant emission on large scales around SR~24N, i.e. the cicrcumbinary disc seen in scattered light and in molecular gas emission does not show a counterpart in ALMA band~6 continuum. This means that either dust growth has been impeded in this area, or the evacuation of evolved dust grains is very efficient.

At this moment it is not possible to infer the dynamical state of SR~24, whether it is on a bound binary orbit or a stellar fly-by. Neither the change of relative separation between the two components has been significantly observable in the past two decades \citep{Schaefer2018}, nor are there any Gaia measurements for relative proper motions. The two stars that constitute the northern source, on the other hand, have been found to be in a bound binary configuration with estimated orbits presented in \citet{Schaefer2018}, where the authors regarded data points from a time span of 25 years. We want to shortly discuss where the estimated locations of the detected continuum discs fall in the astrometrical diagramme and how the new data impacts the estimate of orbital elements.  
\paragraph*{Updating astrometrical measurements of the orbit of SR~24N}
From the separate detection of discs around SR~24Na and SR~24Nb with ALMA we can infer the relative stellar positions by assuming that they are at the centre of the corresponding unresolved continuum emission. 
\begin{table}
\caption{Measurements of relative spacing between SR 24Na and SR 24Nb. The epochs are given in Modified Julian Date (MJD = JD$-$2400000.5). The techniques/instruments are (1) Lunar Occultation, (2) IRTF/NFSCAM, (3) VLT/NACO, (4) Keck/NIRC2, (5) ALMA.}
\label{tab:SR24N}
\begin{tabular}{lllll}
\hline
Epoch & $\rho$ [mas] & PA [deg] & Reference & Technique\\
\hline
48476.5&197&84&\citet{Simon1995}&1\\
52055.0&112&63&\citet{McCabe2006}&2\\
52055.0&120&59&\citet{McCabe2006}&2\\
53126.5&81&45.6&\citet{Correia2006}&3\\
56843.5&93.7&248.0&\citet{Schaefer2018}&4\\
57214.5&99.1&240.7&\citet{Schaefer2018}&4\\
58677.5&116&221.7& This work&5\\
\hline
\end{tabular}
\end{table}
To estimate orbital parameters from the available astrometric data points, we use the publicly available software package {\tt orbitize!}\footnote{\hyperlink{https://orbitize.readthedocs.io/}{orbitize.readthedocs.io}} \citep{Blunt2020}.  The programme returns a selection of fitted orbital parameters to the given data set making use of the Bayesian rejection-sampling method called OFTI \citep{Blunt2017}.
\begin{figure}
    \centering
    \includegraphics[width=1.8\columnwidth]{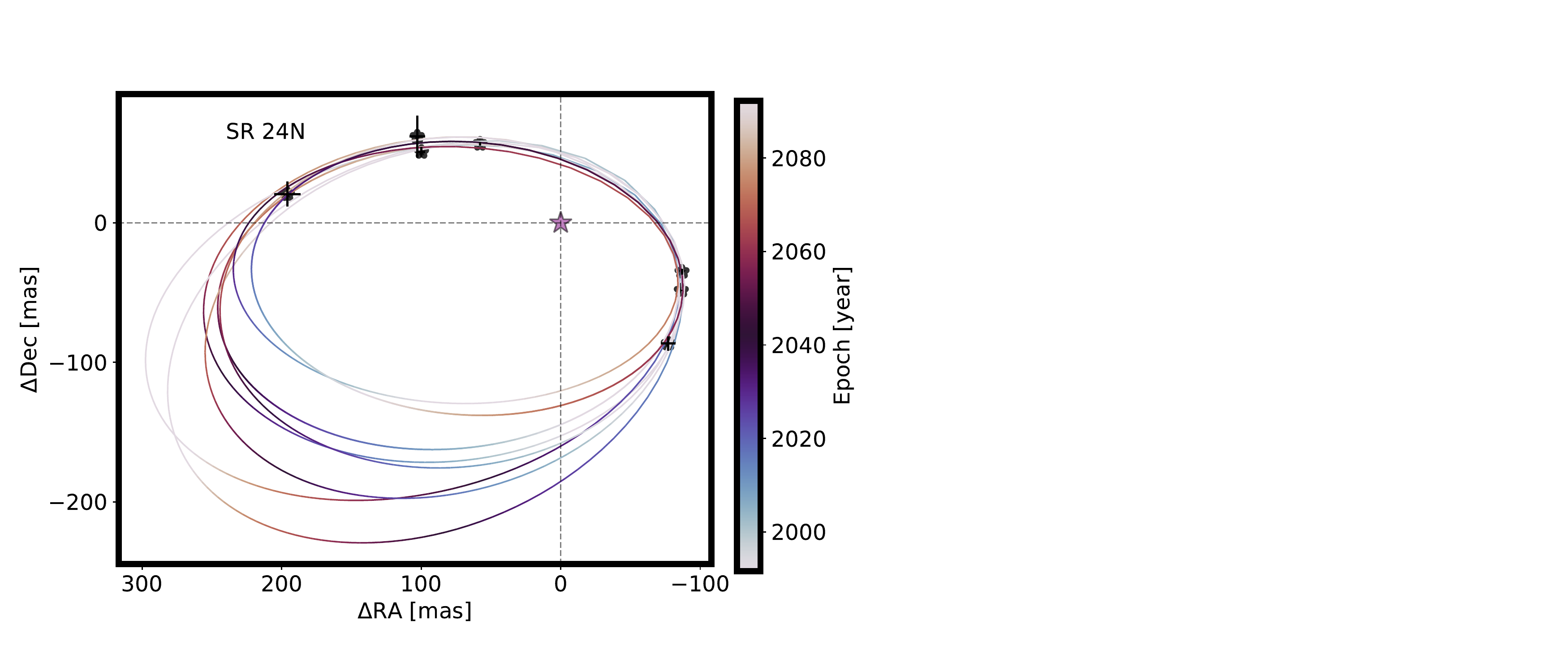}
    \vspace{-0.7cm}
    \caption{Relative positioning of SR~24N binary. The origin is centred on SR~24Na, marked as a purple star, while SR~24Nb is plotted as a black star according to its relative spacing in different epochs, as listed in table~\ref{tab:SR24N}. The solid lines show exemplary orbits of SR~24N calculated with the {\tt orbitize!}-package \citep{Blunt2020}.}
    \label{fig:SR24orbit}
\end{figure}
In Fig.~\ref{fig:SR24orbit} we show the result of ten calculated orbits that fit the data points within their errors. We use these fits for inferring the orbital elements, taking each orbital element's standard deviation within this set of orbits to estimate an error of the calculation. We thus obtain an orbit for the northern binary with a semi-major axis of $a_{\rm N} = (21\pm2)\,$au, an eccentricity of $e_{\rm N}=0.68\pm0.04$, an inclination and position angle of $i_{\rm N}=(131\pm4)^{\circ}$ and $PA_{\rm N}=(257\pm3)^{\circ}$ or $PA_{\rm N}=(77\pm3)^{\circ}$. The ambiguity for the position angle arises because from the projected motion alone, we cannot determine which half of the orbit is near to the observer. The estimated total mass for the northern binary is $M_{\rm N}=(0.53\pm0.04)\,M_\odot$. This results in an orbital period of $P_{\rm N} = (131 \pm 3)\,$yr. All those parameters are within the errors of the values obtained by \citet{Schaefer2018}. 

\paragraph*{Geometrical Model for SR~24S:}
\citet{fernandez-Lopez2017} inspected the gas kinematics of the disc around SR~24S, speculating about the alignment of the disc in space. The authors studied the asymmetry of the gas motion between the east and west of the disc and suggested the near edge of the disc to be located towards the east, the disc rotating counter-clockwise. This assessment relies on the explicit assumption that the gas around SR~24S is accreting.

The SPHERE/IRDIS observation contradicts this perception. The two cone-like shapes opening to the south-east and north-west suggest that the observation captures the top and bottom side of a vertically very inflated disc. In this scenario the brightness asymmetry indicates that the near edge is to the west (i.e. the disc's top half or front side ($FS$) opens towards the south-east). In reference to the discussion in \citet{fernandez-Lopez2017}, this implies that the disc is rotating in clockwise direction and, concluding from their analysis of the kinematics, that the observed molecular ${\rm C}^{18}{\rm O}$(2--1)-line traces expanding rather than accreting gas.

To test whether the scattered light image can be explained by this geometrical setup, we constructed a radiative transfer model using the code RADMC3D\footnote{\hyperlink{https://www.ita.uni-heidelberg.de/~dullemond/software/radmc-3d/}{ita.uni-heidelberg.de/~dullemond/software/radmc-3d/}} \citep{Dullemond2012}. We describe the details of this model in Appendix~\ref{appendix:SR24S_RT}.
\begin{figure*}
    \centering
    \includegraphics{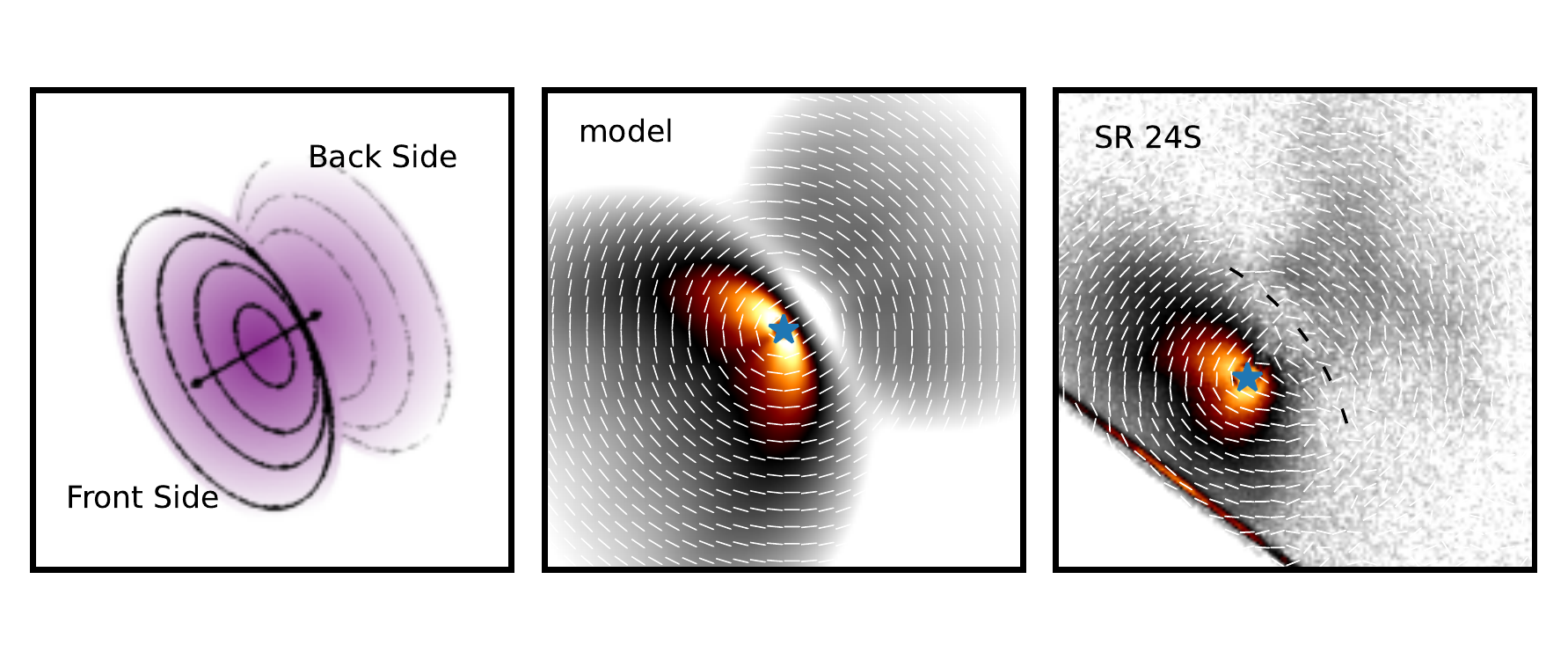}
    \vspace{-1.0cm}
    \caption{Inspecting the geometry of SR~24S: the left image shows a conceptual sketch of the disc, showing the near edge to the west and the `bowl'-shaped scattering surface of the upper disc half opening to the south-east. The central image shows the intensity, overplotted with the $AoLP$, produced by the radiative transfer model (described in Appendix~\ref{appendix:SR24S_RT}) using RADMC3D. The right panel shows a zoom-in onto the SPHERE/IRDIS $H$-band image of SR~24S. The black dashed line indicates the putative mid-plane of the disc.}
    \label{fig:SR24S_RT}
\end{figure*}
The model confirms the notion that the observation can be connected to the two sides of a vertically inflated disc. We further find that the observation of the backside requires a density profile that includes sufficient mass to scatter efficiently at high altitudes, while still being transparent enough such that photons are not absorbed or scattered twice along the line-of-sight. This would not allow the backside photons to escape, as can be seen as vanishing polarised intensity in the inner parts of the backside, both in the radiative transfer model and the observation, marked by a black dashed line in the right panel of Fig.~\ref{fig:SR24S_RT}.

The radiative transfer model futher captures the alignment of the $AoLP$ around the disc's mid-plane, where in both model and observation the vectors form an oval-shaped pattern slightly off-set from the stellar position towards the backside of the disc.

We conclude from this analysis that the asymmetry seen in polarised intensity around SR~24S is likely due to a tilted, flared disc, in contrast to an inner misaligned component as proposed by \citet{Mayama2020} based on HiCIAO data. This is in agreement with gas kinematics shown in \citet{Pinilla2017}. Still we emphasise that neither the molecular lines, nor the SPHERE/IRDIS exclude the possibility of a misaligned inner disc.

\subsubsection{FU~Orionis}
For FU~Orionis we confirm the structures previously observed in \citet{Laws2020} and partly in \citet{Liu2016} and \citet{Takami2018}. The characteristic increase of stellar magnitude has been linked with a prograde, disc-penetrating stellar fly-by \citep{Cuello2020,Borchert2022,Cuello2022}. Analysis of the $^{12}$CO kinematics in \citet{Perez2020} revealed that the northern and southern source have a relative velocity of at least $1\,{\rm km}{\rm s}^{-1}$ along the line-of-sight. The scattered light shows strongly disturbed structures such as the bright extended arm ($B$) east of the northern source, possibly linked to gravitational interaction and a subsequent stellar outburst.
The kinematics of the arm ($B$) link it to both the systematic velocity of the northern and southern source \citep[][]{Perez2020}, further promoting the idea that it results from a recent close stellar encounter.
Interestingly, the ALMA continuum fluxes presented in \citet{Perez2020} show two discs similar in size that appear without substructure at the given resolution (40$\,$mas). Yet, their small sizes might be a result of tidal truncation during a past fly-by event \citep[e.g.][]{Cuello2019}. We also re-detect stripes of reduced intensity ($C$) extending towards the north of the tip of the bright arm and diffuse structures ($D$ and $E$) towards the east of the objects.
\citet{Liu2016} proposed the detection of a further spiral arm in the western region of the primary disc. \citet{Laws2020} did not observe any flux to confirm this suggestion and neither do we detect this features in the observations presented here. Thus, we find that the absence of this arm is not due to the different observational wavelengths ($H$-band and $J$-band), nor due to different epochs.

The primary disc material in FU~OriN seen in the NIR scattered light appears to be considerably more disturbed than in the cases of AS~205 and SR~24. In the fly-by scenario this speaks for a former disc-penetrating passage, disrupting the disc and triggering the FU~Orionis outburst event \citep{Borchert2022}, whereas for AS~205 and SR~24 the interaction appears more moderate, speaking for a passage outside the discs.  

\section{Conclusions}\label{sec:conclusions}
In this work we investigated the polarised scattered light observation of three well known twin-disc systems using the SPHERE/IRDIS instrument of the Very Large Telescope. We outlined how the angle of linear polarisation can be used to expose the dominant light source for a scattering region and investigated the polarised intensity pattern where multiple light sources become relevant. We find that
\begin{itemize}
    \item from the comparison of the measured $AoLP$ to the relative spacing of present light sources one can infer the dominant star for polarised scattering.
    \item polarised intensity is observed as an incoherent sum over different polarisation states produced by scattering of different light sources. In a system of two stars, two nulls of linearly polarised intensity are expected where the incident directions to the two stars are perpendicular and where the linearly polarised intensity due to the light from both star is equal.
\end{itemize}
We used this assessment to constrain that
\begin{itemize}
    \item the bridging area in AS~205 scatters light mostly from the southern component, which is counter-intuitive as AS~205S is fainter than AS~205N by about one order of magnitude.
    \item two polarised intensity nulls appear close to the disc around AS~205N on a circle with the projected locations of the two light sources, as predicted by our polarisation analysis. Whether these decrements are indeed caused by polarisation self-cancellation (due to the two present light sources), or rather by the absence of scattering material (supported by the decrement of $^{12}$CO emission at the location of $N_1$ shown in Fig.~\ref{fig:P_CO_comp_AS205}) remains an open question.
    \item the scattering in the discs in the system of SR~24 is dominated by their respective host star.
    \item the dust in the FU~Orionis system scatters almost exclusively light from the northern source in the NIR. The low relevance of the southern object and stark brightness contrast suggests that it is backgrounded and obscured by significant amounts of fine-grained dust. 
\end{itemize}
Additionally, we compared the SPHERE/IRDIS observation to existing ALMA data. In general, we find vastly smaller disc sizes for mm-sized grains (observed with ALMA) than for $\lesssim 1\,$µm-sized grains (observed with SPHERE). 
For AS~205 we used the $^{12}$CO velocity measurements along the line-of-sight and Gaia measurements in the image plane to infer the binary components' full three-dimensional relative proper motion and discuss the system's geometry with respect to the observer. If the proper motion measurements can be trusted and the system's mass from stellar evolution models is indeed significantly overestimated, as suggested from our results of gas kinematics, the system is most likely representing a stellar fly-by.
For SR~24 we re-investigated the archival 1.3 mm continuum data and detected separate emission from compact circumstellar dust around each component of the northern binary, SR~24Na and SR~24Nb. We used the centre of the detected continuum fluxes to update orbital elements of the northern binary.
The $PI$ asymmetry around SR~24S is likely due to the observation of the front- and backside of an vertically inflated disc.
For FU~Orionis, the scattered light suggests a strong dynamical perturbation of (sub-)µm dust grains. This is in contrast to the restricted, axi-symmetrical, evenly-distributed mm-grains that had been detected in the two discs around the individual binary components \citep{Perez2020}.
\\

Finally, we remark that since many stars exist in multiples, or their formation histories in dense environments suggest stellar encounters, the information on protoplanetary environments in such cases is highly relevant for the discussion of disc evolution and exoplanetary statistics.  

\section*{Acknowledgements}
{We thank the anonymous referee for a careful and constructive report, with a special emphasis on the important hint towards the uncertainties in Gaia measurements. We thank Troels Haugbølle for an interesting discussion on the different mass estimates from stellar evolution models and dynamical measurements.} 
P.W. acknowledges support from FONDECYT grant 3220399 and from ALMA-ANID postdoctoral fellowship 31180050.
S.P. acknowledges support from FONDECYT grant 1191934.  This work was funded by ANID -- Millennium Science Initiative Program -- Center Code NCN2021\_080.
This work has been carried out within the framework of the NCCR PlanetS supported by the Swiss National Science Foundation under grants 51NF40\_182901 and 51NF40\_205606. GG acknowledges the financial support of the SNSF.
N.T.K. and P.P. acknowledge support provided by the Alexander von Humboldt Foundation in the framework of the Sofja Kovalevskaja Award endowed by the Federal Ministry of Education and Research.
A.Z. acknowledges support from the FONDECYT Iniciaci\'on en investigaci\'on project number 11190837.
This project has received funding from the European Research Council (ERC) under the European Union Horizon 2020 research and innovation program (grant agreement No. 101042275, project Stellar-MADE).
L.C. acknowledges support from FONDECYT Grant 1211656.
This work made use of the Puelche cluster hosted at CIRAS/USACH.
The work is based on observations collected at the European Southern Observatory under ESO programmes 098.C-0422(B) and 099.C-0685(A).
This paper makes use of the following ALMA data: ADS/JAO.ALMA\#2016.1.00484.L, ADS/JAO.ALMA\#2018.1.00028.S. ALMA is a partnership of ESO (representing its member states), NSF (USA) and NINS (Japan), together with NRC (Canada), MOST and ASIAA (Taiwan), and KASI (Republic of Korea), in cooperation with the Republic of Chile. The Joint ALMA Observatory is operated by ESO, AUI/NRAO and NAOJ.
This work has made use of data from the European Space Agency (ESA) mission
{\it Gaia} (\url{https://www.cosmos.esa.int/gaia}), processed by the {\it Gaia}
Data Processing and Analysis Consortium (DPAC,
\url{https://www.cosmos.esa.int/web/gaia/dpac/consortium}). Funding for the DPAC
has been provided by national institutions, in particular the institutions
participating in the {\it Gaia} Multilateral Agreement.
\section{Software}
This work has made use of the IRDAP-pipeline \citep[][]{vanHolstein2020} for the processing of SPHERE/IRDIS data, of the {\tt denoise}-package based on the SPLASH tool \citep[][]{Price2007} for noise-reduction, of the {\tt orbitize!}-package \citep[][]{Blunt2020} for orbit fitting to given data points, 
of the {\tt eddy}-package \citep[][]{eddy} to estimate stellar masses and line-of-sight velocities in AS~205,
of RADMC3D \citep[][]{Dullemond2012} for radiative transfer calculations and of the {\tt optool}-package \citep[][]{Dominik2021} for the calculation Müller matrices and opacities.
We further used IPython \citep{ipython}, NumPy \citep{numpy} and Matplotlib \citep{Matplotlib} for data analysis and creating figures.
\section*{Data Availability}
The raw SPHERE/IRDIS data for AS~205 and SR~24 are available under the ESO programme code 099.C-0685(A), FU~Orionis under 098.C-0422(B).
For AS~205 and SR~24, the raw ALMA data is available under the project codes \href{https://almascience.nrao.edu/aq/?project_code/2016.1.00484.L}{2016.1.00484.L} and \href{https://almascience.nrao.edu/aq/?project_code/2018.1.00028.S}{2018.1.00028.S}, respectively. The processed AS~205 data is available as part of the \href{https://almascience.eso.org/almadata/lp/DSHARP/}{DSHARP data release}. The VLT/NACO data is available under the programme IDs 073.C-0121(A), 0103.C-0290(A), 073.C-0379(A)), 073.C-0530(A) and 0101.C-0159(A).


\bibliographystyle{mnras}
\bibliography{example} 



\appendix

\section{{$Q_\phi$ / $U_\phi$ analysis in binary systems}}\label{appendix:QUphi}
{For single star systems, a popular way of presenting the polarised intensity is by showing the $Q_\phi$ and $U_\phi$ images \citep[][]{Schmid2006,deBoer2020}, as defined in equations~(\ref{equ:QUphi}), where $\phi$ is the polar coordinate in the image plane, centred on the star. These fields are constructed such that polarised light originating from the coordinate centre and scattered only once contributes exclusively a positive value to $Q_\phi$. Consequently, $U_\phi$ is typically inspected to reveal image regions where those assumptions break down \citep{Canovas2015,Pohl2017}. In multiple system, where the projected distance between different stars is comparable to relevant image scales, the condition of light only originating from the centre is trivially violated as there are one or more off-centred light sources in the system. It is thus not possible to interpret $Q_\phi$ and $U_\phi$ in the same manner.} 
\begin{figure*}
    \centering
    \begin{subfigure}{1.0\textwidth}
    \includegraphics{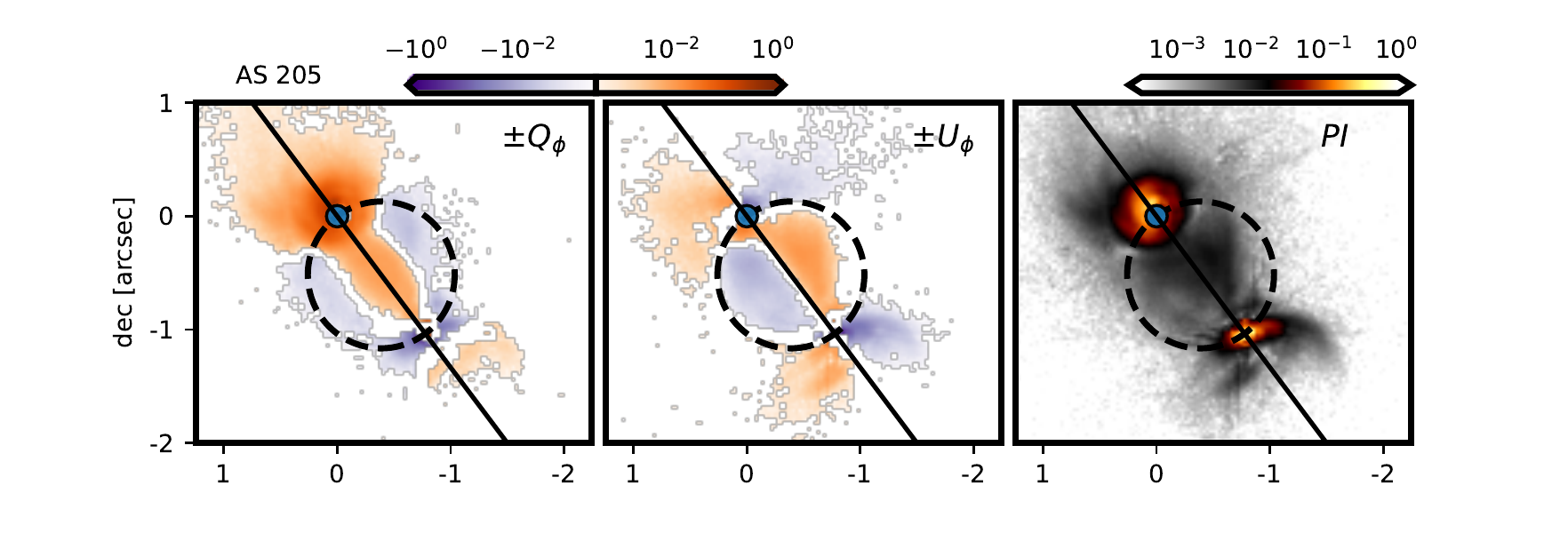}
    \vspace{-1.5cm}
    \end{subfigure}
    \begin{subfigure}{1.0\textwidth}
    \includegraphics{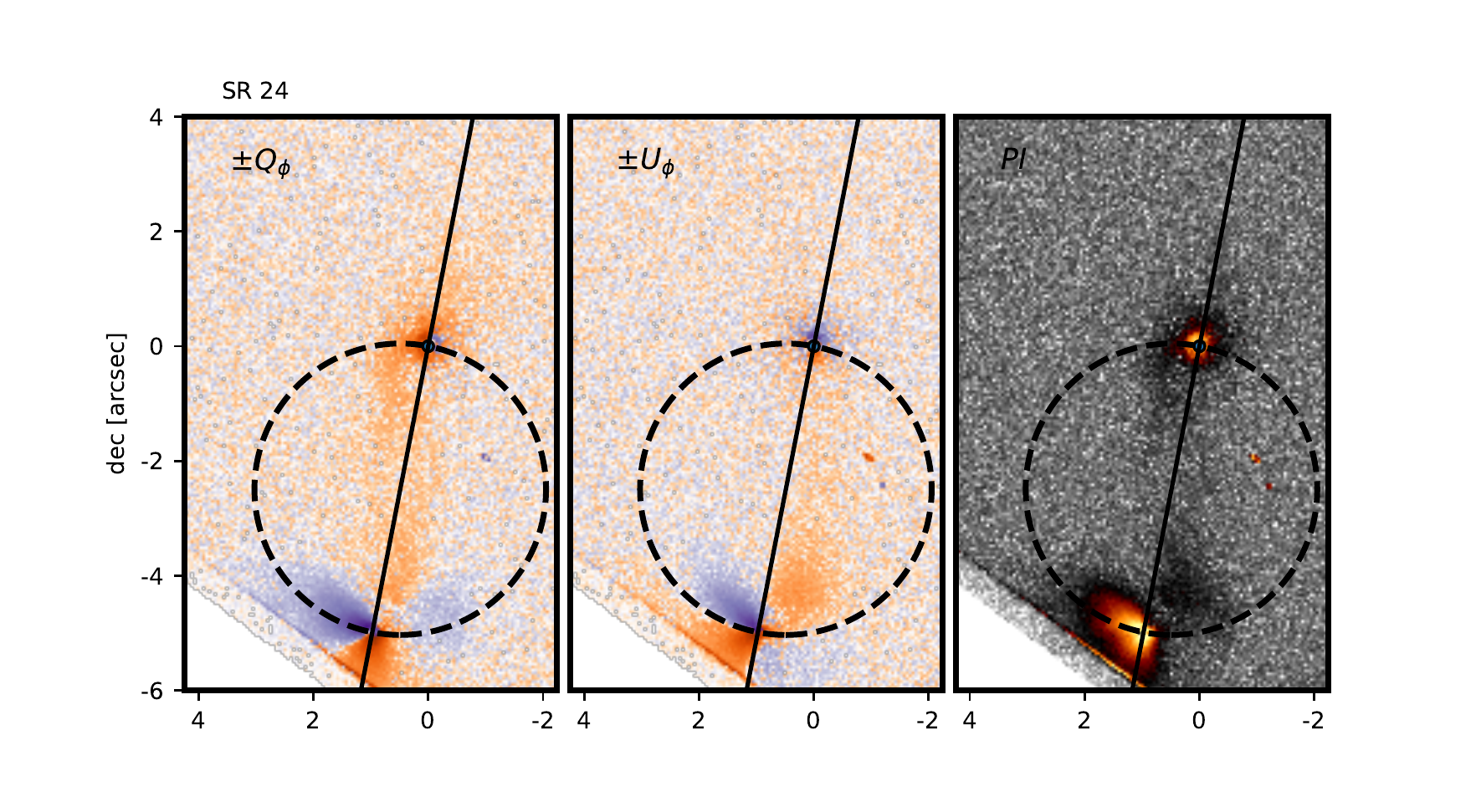}
    \vspace{-1.5cm}
    \end{subfigure}
    \begin{subfigure}{1.0\textwidth}
    \includegraphics{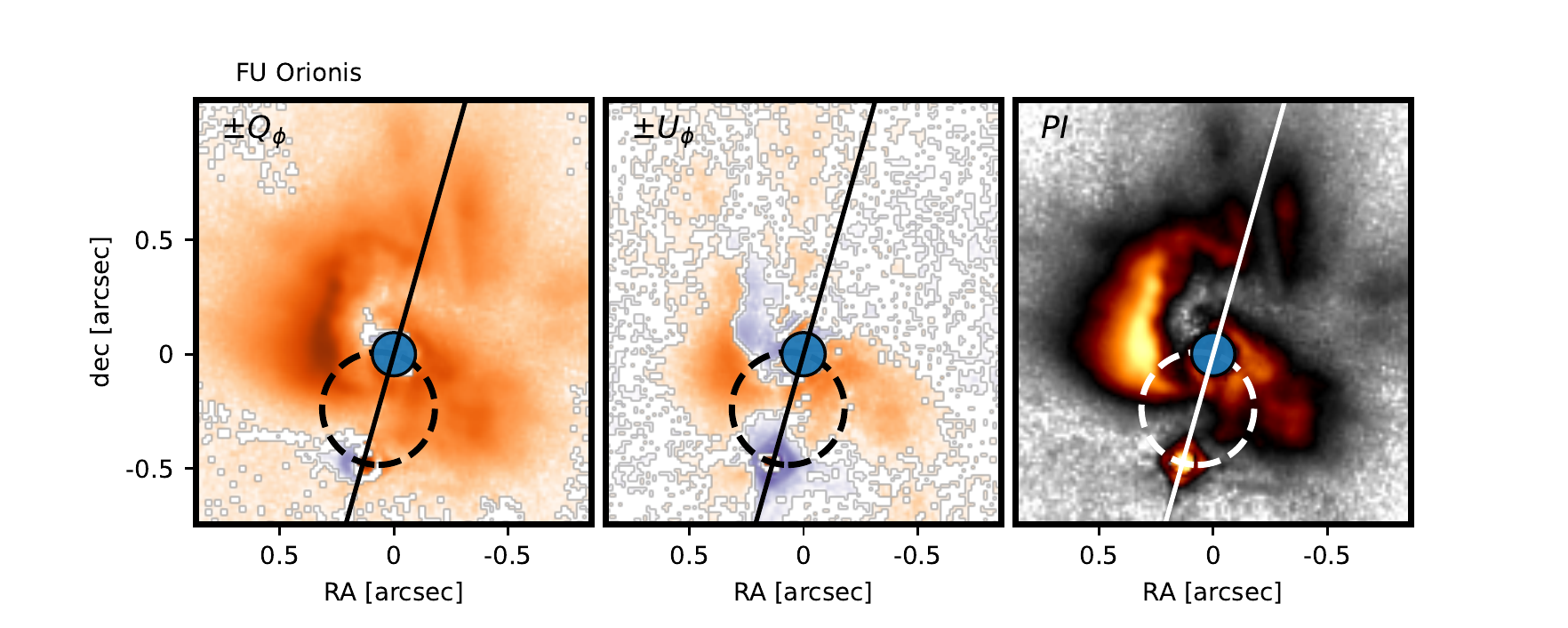}
    \end{subfigure}
    \caption{{$Q_\phi$, $U_\phi$ and $PI$ images for the three systems under investigation, AS~205, SR~24 and FU~Orionis. The $Q_\phi$ and $U_\phi$ colour maps show both the positive (orange) and negative (purple) intensities on logarithmic scale. In all the images the black line and the dashed circle go through the two stars. For image locations along the black line, the incident directions are the same towards both stars (or shifted by 180$^\circ$). For image locations along the dashed circle, the incident directions towards the stars are perpendicular. In the bottom right panel the black colour of line and circle were replaced by white colour for visibility.}}
    \label{fig:QUphi}
\end{figure*}

{Still, the images shown in Fig.~\ref{fig:QUphi} hold some interesting information. We show there the $Q_\phi$ and $U_\phi$ images on logarithmic scale, separated into their positive and negative components. Along with this, we show again the total linearly polarised light, that combines those fields, since $PI = \left(Q_\phi^2+U_\phi\right)^{0.5}$, which can be derived by combining equations~(\ref{equ:PI}) and (\ref{equ:QUphi}). As in a single star system, the single-scattering events from the star at the coordinate centre contribute to positive $Q_\phi$.
We construct a straight line connecting the two stellar components for each system. Along this line the incident directions in the image plane towards the two stars are identical or rotated by 180$^\circ$ towards each other, i.e. the expected $AoLP$ for single-scattering events in these regions is identical and hence, along this line, also the second star adds exclusively positively to $Q_\phi$.}

{We further construct a circle defined by the distance between the stars. As shown in Sec.~\ref{sec:two_light_sources}, along this circle the incident directions for light of the two stars are perpendicular in the image plane. Thus, by construction, along this circle the scattered light from the off-centre star contributes exclusively negatively to $Q_\phi$. Consequently, $Q_\phi$ becomes negative, where the observed scattered light originates predominantly from the southern star. This is the case for most parts in AS~205. For SR~24 the signal-to-noise ratio is not large enough between the two stars to allow any robust conclusions. In FU~Orionis, almost all the image is dominated by the positive $Q_\phi$ component, again pointing to the dominance of light scattering from FU~OriN.}

{We can further see that in all three systems $U_\phi$ contains relevant intensity, mainly due to the additional off-centred light source. Lets define this light source to be located at an azimuthal angle of $\phi_{\rm B}$ towards the centre of the frame ($\phi$ increasing in anti-clockwise direction). Inside the dashed circles in Fig.~\ref{fig:QUphi}, single-scattering from the second light source is expected to contribute negatively where $\phi<\phi_{\rm B}$ and positively where $\phi>\phi_{\rm B}$, and vice versa outside the circle. For AS~205, the $U_\phi$ image shows this expected pattern almost globally, even to the north of AS~205N. This suggests that some parts of the scattered polarised light that we see from structures north of AS~205N are caused by scattering light from the southern source. 
There is one exception of a small area close the AS~205N, to its south, where the $U_\phi$ signal is positive even though located in the negative half of the circle. The second light source cannot account for this contribution, it might be due to effects previously described in single-star systems \citep[][]{Canovas2015,Pohl2017}.}

{For FU~Orionis the $U_\phi$ image shows an interesting structure. There is strong signal in large parts of the image, especially in the bright arm (marked as $B$ in Fig.~\ref{fig:mosaic}), where $U_\phi$ flips from positive to negative. In general, the image does not suggest a strong contribution from the secondary light source, but rather that the intricate environment contributes relevant multiple scattering events.}

\section{Subtracting the stellar polarisation in binary systems}\label{appendix:star_pol}
\begin{figure*}
    \centering
    \includegraphics{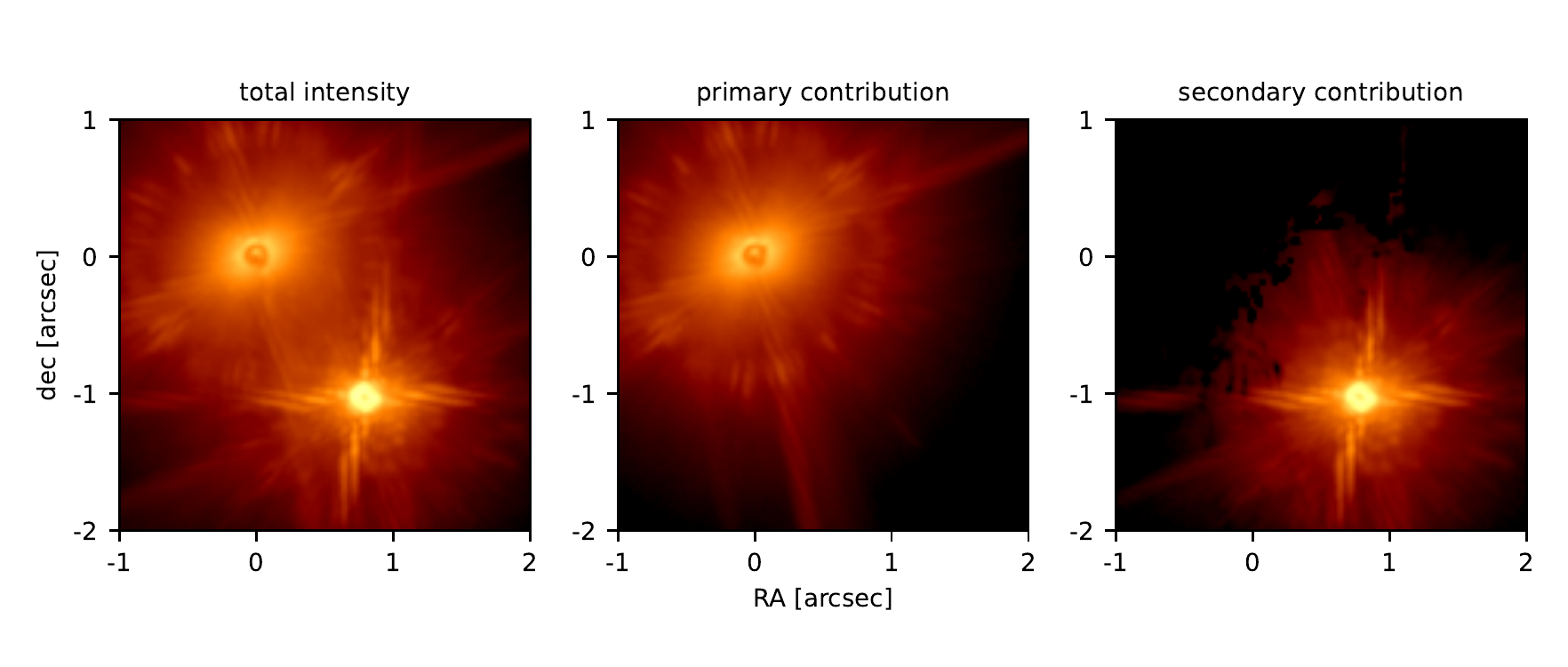}
    \vspace{-1cm}
    \caption{Disentangling of stellar intensity contributions. Left image shows original observed intensity after reduction with IRDAP. The central panel shows the extracted contribution associated with AS~205N, the right panel shows the contribution associated with AS~205S.}
    \label{fig:star_disentang}
\end{figure*}

\begin{figure*}
    \centering
    \includegraphics{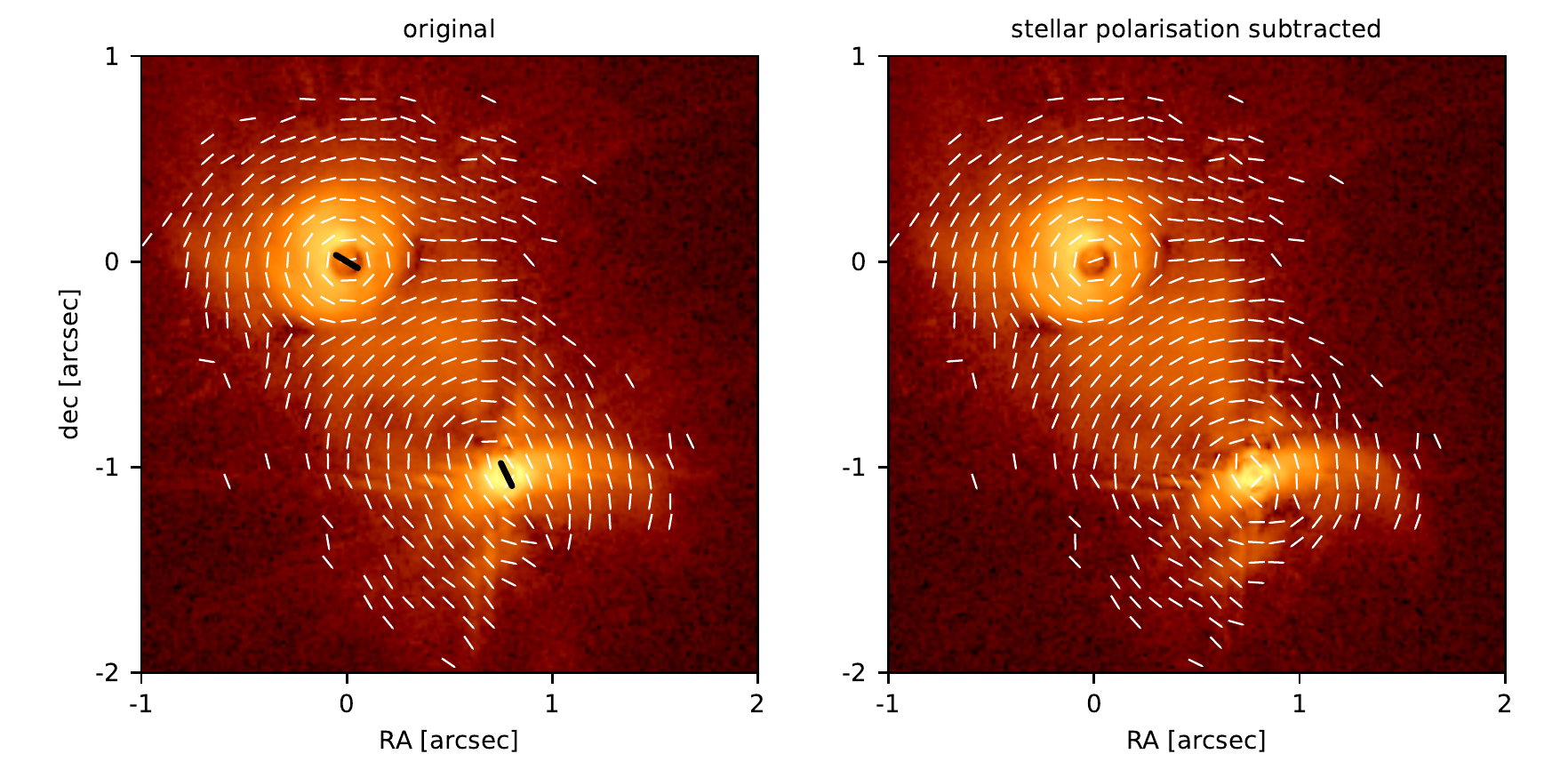}
    \vspace{-0.5cm}
    \caption{AS~205 before (left) and after (right) subtracting the estimated stellar polarisation. The overplotted white vectors indicate the $AoLP$. As a reminder: the measured $AoLP$ of AS~205N is $\sim59^\circ$, and for AS~205S it is $\sim 26^\circ$ (measured east-of-north and visualised as a black bar in the left panel). Especially the presence of the southern contribution is clearly visible in the $AoLP$ map in the left panel.}
    \label{fig:subtr_comparison}
\end{figure*}

Direct stellar emission is unpolarised in broadband filters. Yet, the presence of dusty material surrounding the star on scales smaller than the telescope resolution, or within the observer's line-of-sight, can add a $DoLP$ to the stellar light within the telescopes star-centred point spread function (PSF). This means that the stellar diffraction features in the image are also slightly polarised and thus alter the images of $PI$ \citep{Canovas2011,vanHolstein2020}. 
Whether a star's halo increases or decreases the local signal depends on the location in the disc and the $AoLP$ of the stellar polarisation.
This is problematic especially in regions where the stellar intensity is high and especially for a precise calculation of the $AoLP$, as the stellar polarisation super-imposes a directional off-set. To isolate the polarisation that is due to local scattering, we therefore aim to subtract the stellar polarisation and the constant polarised background.
We measure the background polarisation in selected patches in the image that are free seemingly free of intensity above noise level, both in polarised and total intensity. The seleceted areas are listed in Table~\ref{tab:masks}.
In a system of a single dominant light source, one way to infer the stellar polarisation is by measuring separately the polarisation components $Q$ and $U$, and the corresponding total intensity components $I_Q$ and $I_U$, in regions of the stellar halo where there is no signal due to the circumstellar material. There, the stellar halo is isolated such that the fractional components can be attributed to unresolved stellar polarisation, $q_\ast=Q/I_Q$ and $u_\ast=U/I_U$. One can then impose this fraction onto the entire intensity field, to subtract this part from the observed $Q$ and $U$ components. This gives the components corrected for the stellar contributions:
\begin{eqnarray}
    Q_{\rm cor} &=& Q - q_\ast \times I_Q\,,\label{equ:app_star_pol1}\\
    U_{\rm cor} &=& U - u_\ast \times I_U\,.\label{equ:app_star_pol2} 
\end{eqnarray}
In a binary system we find two challenges: first of all, the scattered light is mostly extended and asymmetric, which makes it hard to find a region in the images that is devoid of circumstellar material.  
Secondly, equations~(\ref{equ:app_star_pol1}) and (\ref{equ:app_star_pol2}) fail if the fractional stellar polarisation is not constant throughout the image. This is definitely the case in a system of multiple stars with different polarisation states. To adapt the method, we aim to subtract the intensity contribution of the individual stars separately:
\begin{equation}\label{equ:Q_cor}
    Q_{\rm cor} = Q - q_{\ast1}\times I_{Q1}- q_{\ast2}\times I_{Q2}\,.
\end{equation}
and equivalently for $U$. Next to a precise measurement of the stellar polarisation, it is important to disentangle the intensity contributions of the two stars in the system, especially where they overlap. To do so, we assume the stars' intensity to be point-symmetric. We start by extracting the northern source.
\begin{table}
\begin{center}
\caption{{Masks employed for measuring background ($\blacksquare$) and stellar ($\star$) polarisations.}}
\begin{tabular}{ |c||c|c|c| }\label{tab:masks}
 {\it Object} & $\Delta R ["]$ & $\Delta \theta_1[^\circ]$& $\Delta \theta_2[^\circ]$\\
 \hline
 \hline
 {\it AS~205 ($\blacksquare$)} & 4.41\,--\,5.15 & -20\,--\,90 & 160\,--\,270 \\
  {\it AS~205N ($\star$)} & 0.735\,--\,1.164 & 20\,--\,100 & 190\,--\,270 \\
 {\it AS~205S ($\star$)} & 0.735\,--\,1.164 & 185\,--\,245 & -95\,--\,50 \\
 \hline
 {\it SR~24 ($\blacksquare$)} & 4.41\,--\,5.15 & -60\,--\,60 & 120\,--\,140 \\
 {\it SR~24N ($\star$)} & 0\,--\,0.18 & 0\,--\,360 &  \\
 {\it SR~24S ($\star$)} & 0\,--\,0.18 & 0\,--\,360 &  \\
 \hline
 {\it FU~Ori ($\blacksquare$)} & 4.41\,--\,5.15 & -20\,--\,90 & 160\,--\,270 \\
{\it FU~OriN ($\star$)} & 1.12\,--\,1.59 & 30\,--\,60 & 100\,--\,180 \\ 
{\it FU~OriS ($\star$)} & 0.15\,--\,0.25 & 0\,--\,360 & \\ 
 \hline
 \hline
\end{tabular}
\end{center}
      \small
      {\bf Notes.} {The masks are annulus segments. All background polarisation masks are centred on the centre of the frames (northern star), all stellar masks are centred on the respective star location in the image. $\Delta R$ provides the inner and outer radius of the annulus, $\Delta \theta$ the starting and end angle of the segment with 0$^\circ$ pointing to the west and positive angles rotating counterclockwise.}
\end{table}
The intensities of the two stars mostly only overlap in the area between them. We, therefore, select a semicircle directed opposing to the binary companion (where the northern component's signal is expected to be mostly uncontaminated) and copy the intensity from there into the other half of the image, respecting point-symmetry towards the star. This is taken as the intensity field of the northern source. We subtract this intensity from the total intensity, which leaves the contribution from the southern star and minor contribution from the inter-binary dust. To correct for this, we repeat the procedure of point-symmetry for the southern source. We show the resulting individual stellar intensities for the example of AS~205 in Fig.~\ref{fig:star_disentang}. 
After separating the stellar intensities, we measure the stellar polarisations in masks listed in Table~\ref{tab:masks} and we use equation~\ref{equ:Q_cor} to correct for the stellar polarisation. The comparison of pre- and post subtraction of the stellar polarisation is shown in Fig.~\ref{fig:subtr_comparison} for the example of AS~205.
While the $PI$ changes only slightly, we note that especially the $AoLP$ around the southern source shows a centro-symmetric pattern after the subtraction. The southern source has a notable level of stellar polarisation, the $AoLP$ in the original image is, therefore, strongly bent towards the $AoLP$ of the star ($\sim26^\circ$).
This simultaneously means that also the $PI$ is strongly affected by the stellar polarisation in the original image close to the secondary. 

\section{{Uncorrected and de-noised image of SR~24}}\label{appendix:SR24_0}
\begin{figure*}
    \centering
    \includegraphics{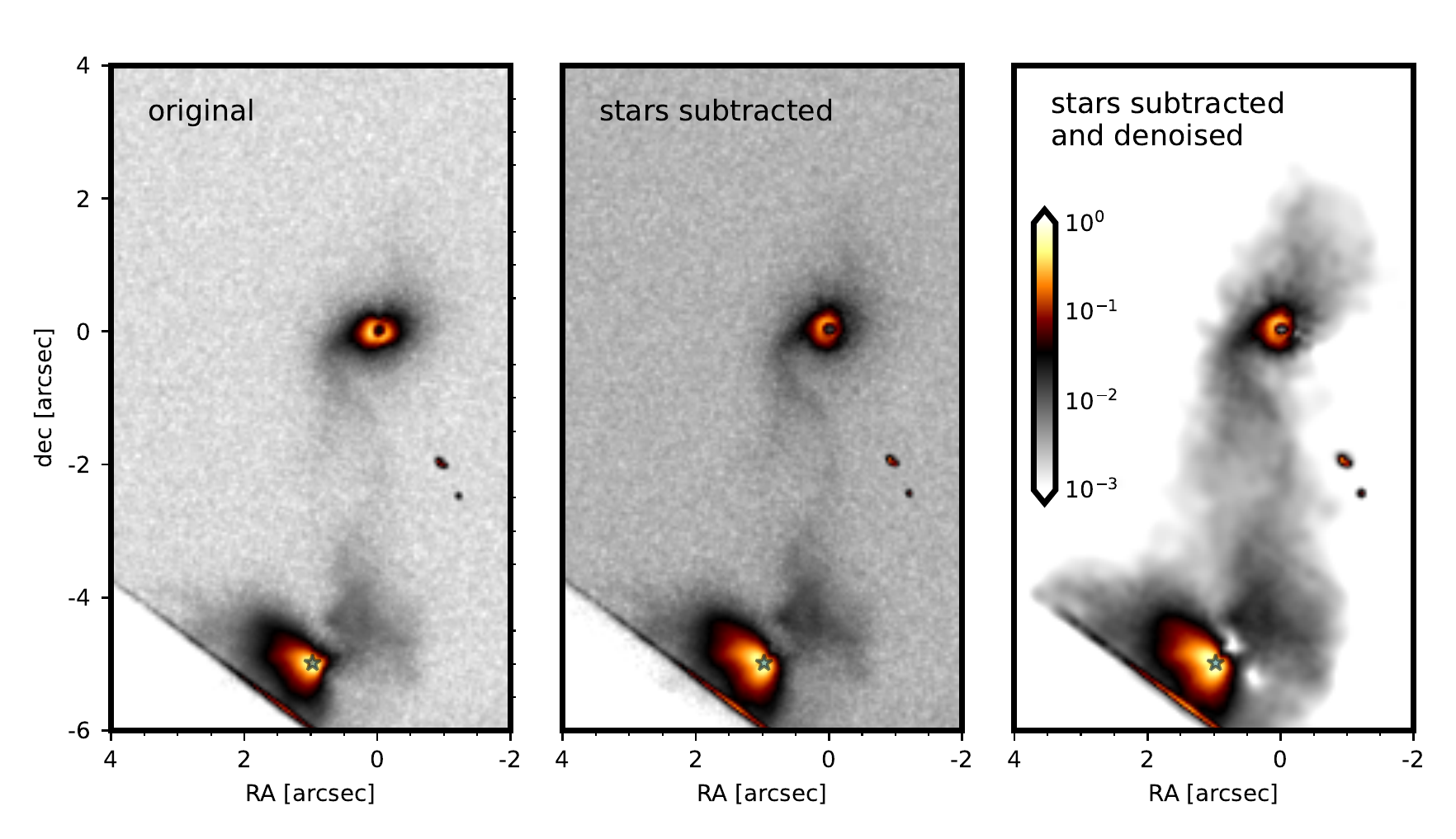}
    \vspace{-0.8cm}
    \caption{All intensities are normalised to their respective maximum value and represented by the same colour bar. The left panel shows the polarised intensity for SR~24 as the direct product of the IRDAP pipeline. The central image shows the polarised intensity after correcting for unresolved polarisation. The right image shows the polarised intensity after additionally applying the {\tt denoise} package with {\tt --beam=1.5} that reduces the Poisson noise.}
    \label{fig:SR24_0}
\end{figure*}

After applying the IRDAP pipeline to the raw frames of SR~24 we applied the subtraction of the estimated unresolved polarisation as described in Sec.~\ref{subsec:SR24_unresolved_pol}.
The correction for unresolved polarisation mainly relies on a confidential measurement of the stellar polarisation, without much contamination from surrounding material. This is very challenging in the case of SR~24, which is why we measured the $DoLP$ and $AoLP$ for both stars with a high level of uncertainty. Therefore, we show the direct IRDAP product in Fig.~\ref{fig:SR24_0}, where the unresolved polarisation has been unaccounted for. The central panel of Fig.~\ref{fig:SR24_0} then shows the image after reduction of the estimated stellar polarisation for comparison. 

{The observation of SR~24 in NIR scattered light shows rather low signal-to-noise, especially in the bridge area ($B$). For this reason, we additionally reduce the Poisson noise of the $Q$ and $U$ products by applying the {\tt denoise}-package\footnote{\label{footnote:denoise}\href{https://github.com/danieljprice/denoise}{github.com/danieljprice/denoise}} (option {\tt -{}-beam=1.5}) which makes use of the particle smoothing kernel of the visualisation tool {\tt SPLASH} \citep[][]{Price2007}. As shown in \citet[][]{Menard2020}, the de-noising procedure yields best results when applied to the $Q$ and $U$ products before combining them to the total linearly polarised image. The reason for this is that $Q$ and $U$ are signed quantities, i.e. the average of pure random noise in the $Q$ and $U$ fields is zero. $PI$ on the other hand is strictly positive, such that the average of random noise still yields a non-vanishing signal.}

{Yet, we warn that in regions of a strong variability of the $AoLP$ the ${\tt denoise}$-package may introduce artificial polarisation cancellation as the $Q$ and $U$ components are smoothed with an adaptable kernel. The benefit of applying this technique is that we decrease the noise level of the image by a factor of $\sim 20$.}
The resulting image is shown in the right panel of Fig.~\ref{fig:SR24_0}, where we can see that structures around and between the stellar components become more prominent as the noise is smoothed. Especially the bridge area between AS~205N and AS~205S is more exposed in this version. 

\section{Radiative Transfer Model for SR~24S}\label{appendix:SR24S_RT}
Here, we describe the setup of the radiative transfer model we performed to qualitatively reproduce the scattered light image shown in Fig.~\ref{fig:SR24S_RT}. We use RADMC3D with the {\tt scattering\_mode\_max=5} option to account for fully anisotropical scattering.
The model employs a spherical grid with a star at its centre (effective temperature of $4170\,$K and radius of $2.7\,{\rm M}_\odot$ \citep[][]{vanderMarel2015}, and spans radially from 5$\,$au to $150\,$au.  
We set up a disc of small dust grains that are expected to be responsible for scattering and extinction at $\lambda=1.625\,$µm.
Therefore, we use a maximum grain size of $a_{\rm max}=1\,$µm and a dust-size distribution of $n(a)\propto a^{-3.5}$. 
We use the {\tt optool}-package \citep{Dominik2021} to calculate the absorption and scattering opacities and Müller matrix elements, assuming the material to follow the DIANA standard model \citep[][]{Woitke2016}.
The density is calculated in every cell:
\begin{equation}
    \rho(R,z) = \frac{\Sigma}{\sqrt{2\pi} h R} \exp \left(-\frac{z^2}{2(h R)^2}\right)\,,
\end{equation}
where $R$ and $z$ are cylindrical coordinates that we transform to spherical coordinates to implement this analytical formula. We assume that both the surface density, $\Sigma$, and the aspect ratio, $h$, follow a power-law profile of $R$. We tested several combinations for both profiles with a number of $10^{7}$ photon packages and found reasonable agreement to the observation for 
\begin{equation}
    \Sigma = 1.2\times10^{-3}\,{\rm g}{\rm cm}^{-2} \left(\frac{R}{10\,{\rm au}}\right)^{-1.0}\,,
\end{equation}
and 
\begin{equation}
    h = 0.15 \left(\frac{R}{10\,{\rm au}}\right)^{1.0}\,.
\end{equation}

\bsp	
\label{lastpage}
\end{document}